\documentclass[aps,pre,preprint,groupedaddress,showpacs]{revtex4-1}

\usepackage{hyperref,graphicx,amsmath,subfig,float}

\begin{document}

\title{Lipophilic Force Driven Dynamics in Langmuir monolayers: In-plane Coalescence and Out-of-plane Diffusion}
\author{U K Basak}
\email{uttam.basak@saha.ac.in}
\author{A Datta}
\email{alokmay.datta@saha.ac.in}
\affiliation{Saha Institute of Nuclear Physics, Surface Physics and Material Science Division, 1/AF Bidhannagar, Kolkata-700064}

\date{\today}

\begin{abstract}
While monolayer area fraction versus time ($A_{n}-t$) curves obtained from surface pressure-area ($\pi-A$) isotherms for desorption-dominated (DD) processes in Langmuir monolayers of fatty acids represent continuous loss, those from Brewster Angle Microscopy (BAM) also show a 2D coalescence. For nucleation-dominated (ND) processes both techniques suggest competing processes, with BAM showing 2D coalescence alongside multilayer formation. $\pi$ enhances both DD and ND with a lower cut-off for ND, while temperature has a lower cut-off for DD but negligible effect on ND. Hydrocarbon chain length has the strongest effect, causing a cross-over from DD to ND dynamics. Imaging Ellipsometry (IE) of horizontally transferred films onto Si(100) shows Stranski-Krastanov (SK) like growth for ND process in arachidic acid monolayer resulting in succesive stages of monolayer, trilayer, multilayer islands, ridges from lateral island-coalescence and shallow wavelike structures from ridge-coalescence on the film surface. These studies show that lipophilic attraction between hydrocarbon chains is the driving force at all stages of long term monolayer dynamics.
\end{abstract}

\maketitle

\section{Introduction}
Langmuir monolayers (LM), composed of amphiphilic molecules, have a wide range of applications especially in mimicking biological membranes and in growing Langmuir-Blodgett (LB) multilayers of tunable thickness and packing density \cite{brezesinski2003,alegria1991,thakur2009,okahata1989,nunes2011,torrano2013,schulte2013,serro2014,thakur2009,tovani2013} with applications in, for example, electrical, electronic and optical device fabrication\cite{ashwell1999,krstic1998,roberts1983,sugi1985,tieke1990,vincett1980}. Stability of LMs is essential for studying their physico-chemical properties as well as ensuring the perfection and thus the reproducibility of the LB multilayers grown from the LM. Langmuir monolayers are at various levels of metastability above the equilibrium spreading pressure (ESP) i.e. the surface pressure ($\pi = \gamma -\gamma_{0}$, $\gamma_{0} (\gamma)$ being surface tension of pure (monolayer-covered) water) spontaneously generated when the bulk amphiphile is brought in contact with a water surface \cite{iwahashi1985, siegel1992} and they destabilize through 2D to 3D transition, turning into bilayers or multilayers. When these multilayer grow in air the process is called `nucleation' \cite{vollhardt1996} while the movement of molecules from the monolayer to water is called `desorption'\cite{smith1980, diamant2000}. Both processes are irreversible for single-tailed amphiphiles.

In this system there are three short-range molecular forces - the hydrophilic attraction between headgroups of amphiphilic molecules and water, the hydrophobic repulsion between the tails and water, and the lipophilic attraction between the tails of adjacent molecules. During the formation of monolayers the competition between the first two forces plays the key role while the importance of the third grows as the surface density is increased, as is expected. Growth of Langmuir monolayers of fatty acids with emergence of different structural phases at different surface concentrations or surface pressures have been thoroughly studied\cite{kaganer1999} as have been the dynamics at very high $\pi$-values, i.e.,collapse\cite{kundu2006, kundu2005, lee2008, ybert2002, baoukina2008, birdi1994}. However, even basic questions regarding the long-term dynamics of Langmuir monolayers at lower surface pressure in the purported `stable' zones, such as the specific differences in dynamics of monolayers that destabilize through a desorption-dominated (DD) mechanism from those undergoing a nucleation-dominated (ND) destabilization have not been addressed and the major destabilizing force among the above three has not been identified. This is essential in order to understand and control the process of destabilization and requires (a) long-term study of the dynamics of monolayer under the DD and ND processes, i.e., with amphiphiles having different tail-lengths, (b) combination of probes at different length scales, and (c) probing the dynamics through field parameters like $\pi$ and temperature.

In this communication, the long-term in-plane dynamics in Langmuir monolayers of single-tailed amphiphilic fatty-acids at different temperatures and surface pressures away from collapse pressure are studied macroscopically through Surface Pressure - Specific Molecular Area ($\pi-A$) isotherms and mesoscopically through Brewster Angle Microscopy (BAM). We have considered the tail-length of the amphiphilic fatty acid molecule to be an `internal' parameter for the destabilization dynamics and have studied its effect through data on a series of amphiphilic fatty acids with tail lengths varying from 14 to 20 carbon atoms. We have also studied the out-of-plane dynamics of a long-chain fatty acid through Imaging Ellipsometry (IE).

\section{Experimental Details}
Amphiphilic fatty acids Myristic acid (C14), Palmitic acid (C16), Stearic Acid (C18), and Arachidic acid (C20), containing the same polar carboxylic (COOH) headgroup but 14, 16, 18, and 20 carbon atoms in their tails, respectively, with quoted purity $>$ 99\% (Sigma-Aldrich) were dissolved in Chloroform (Merck) to prepare 3 mM solutions, spread in a KSV-NIMA Langmuir trough on Milli-Q water (resistivity 18.2 M$\Omega$-cm) at room temperature (25$^\circ$C) and compressed with a speed of 5 cm$^2$/min after solvent evaporation and equilibration.

 Surface pressure was measured by a Pt Wilhelmy plate. Relaxation curves were obtained by recording monolayer area with time at constant surface pressures of $\pi =$ 10 mN/m to 40 mN/m (at 5 mN/m intervals) for C20 and C18, at $\pi=1$ mN/m and from 5 mN/m to 35 mN/m at 5 mN/m intervals for C16, and at $\pi =$ 1 mN/m, 2 mN/m and from 5 mN/m to 25 mN/m at 5 mN/m intervals for C14. Data was collected at 10$^\circ$C, 15$^\circ$C, 20$^\circ$C and 25$^\circ$C by  maintaining subphase temperature using Julabo Recirculating Cooler (FL300).
 
  Brewster Angle Microscopy of monolayers was performed by an Imaging Ellipsometer (Accurion GmBH) in the BAM mode. Laser intensity was kept high to increase the contrast of the BAM image and thereby distinguish between monolayers and multilayers. Time for a monolayer to transform entirely into multilayers was obtained from BAM movie filmed at 8 frames per second (fps) during monolayer relaxation via nucleation. To obtain ellipsometry thickness map of C20 monolayers, all the films are deposited on a hydrophilic Si(100) substrate at different relaxation times at 25 mN/m, 25$^\circ$C using Modified Langmuir-Schaefer (MILS) technique\cite{kato1994,lipp1998,mukherjee2009}. Ellipsometric measurements were performed on the deposited samples using an EP3 imaging ellipsometer (Accurion, GmbH, Germany). For layers appreciably thinner than the wavelength of the probing light the $\Delta$ value is sensitive to the change in the layer thickness, while the $\Psi$ value is hardly affected. We obtain the ellisometric angle $\Delta$ map for each sample. The optical modeling was performed with the software EP4Model (Accurion) to obtain the thickness map of the sample. The details of Imaging Ellipsometry and Modeling can be found \cite{nielsen2013}. The optical functions of Crystalline Si and SiO$_2$ are well-known and implemented in the EP4Model software. For Arachidic acid, we use a single fixed value of refractive index (n=1.457) in all the thickness maps. Absorption by such a thin film is neglected (k=0).

\section{Desorption Dynamics}
\begin{figure}
\centering
\subfloat[]{\includegraphics[scale=0.5]{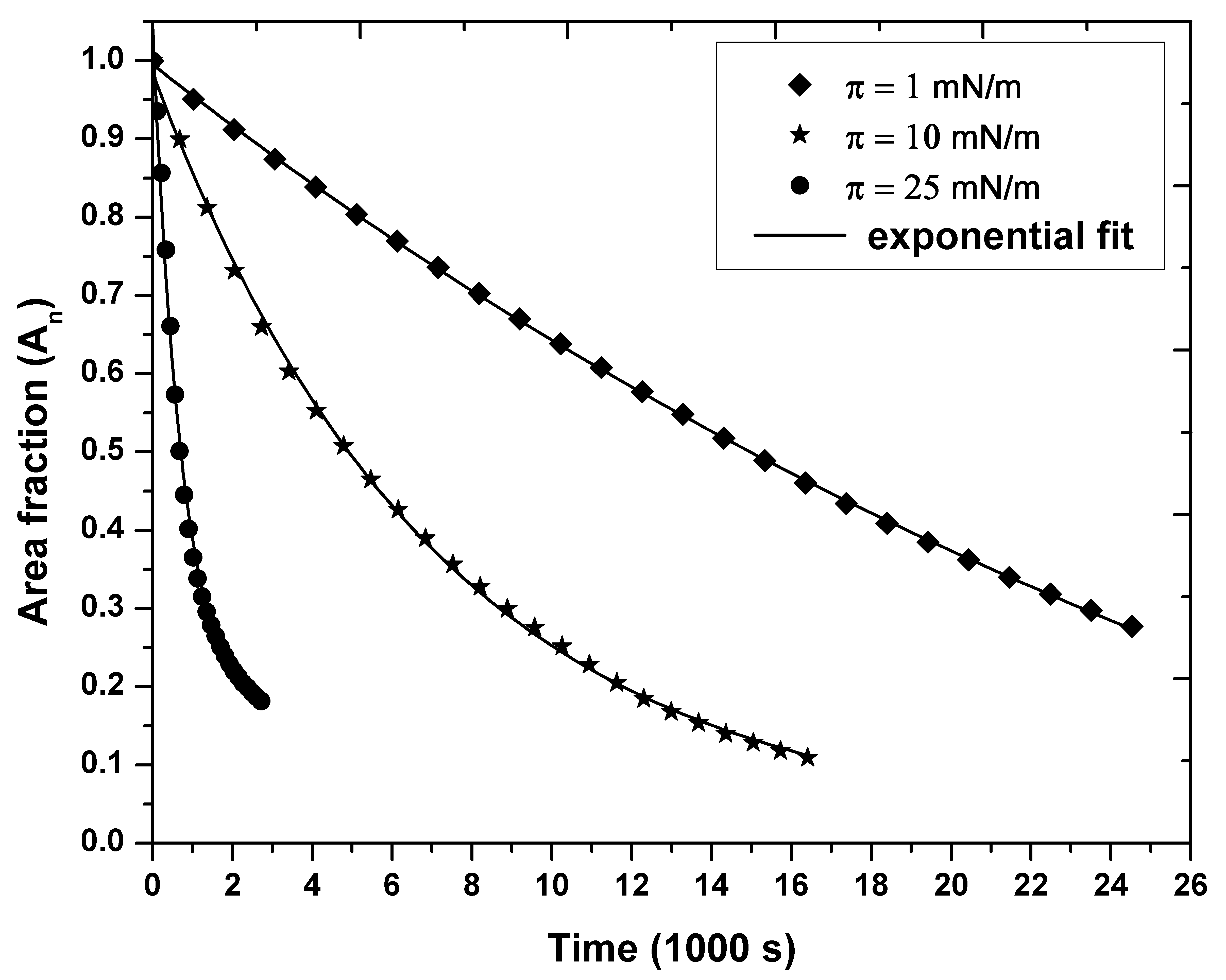}\label{maat}} \qquad
\subfloat[time $\sim$ 1 min]{\includegraphics[width = 1.9 in]{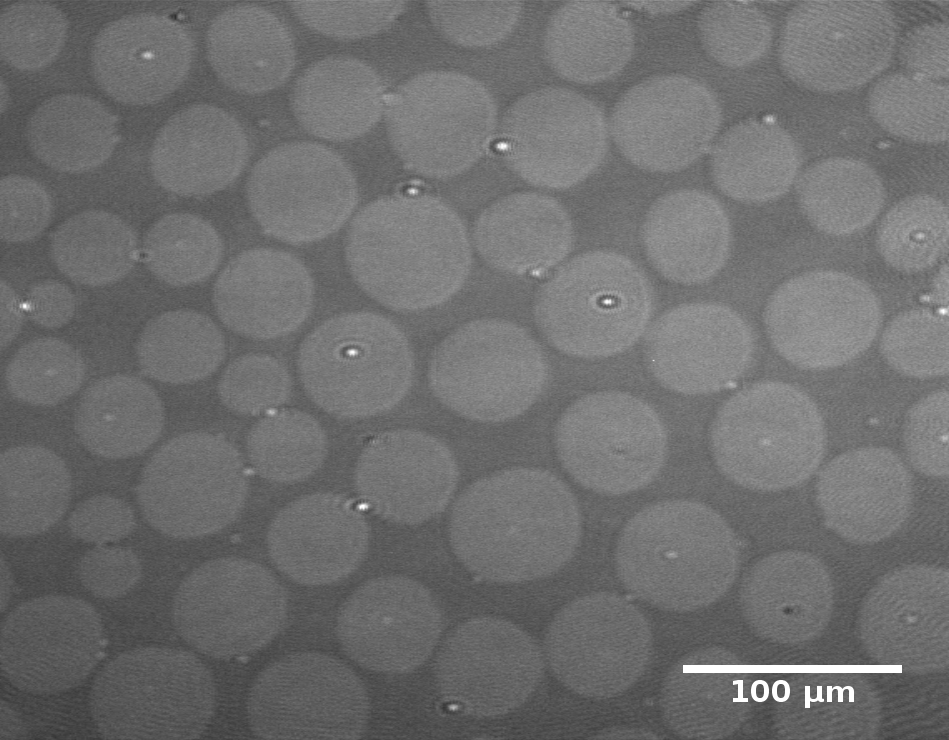}\label{mat1}} \\
\subfloat[time $\sim$ 40 min]{\includegraphics[width = 1.9 in]{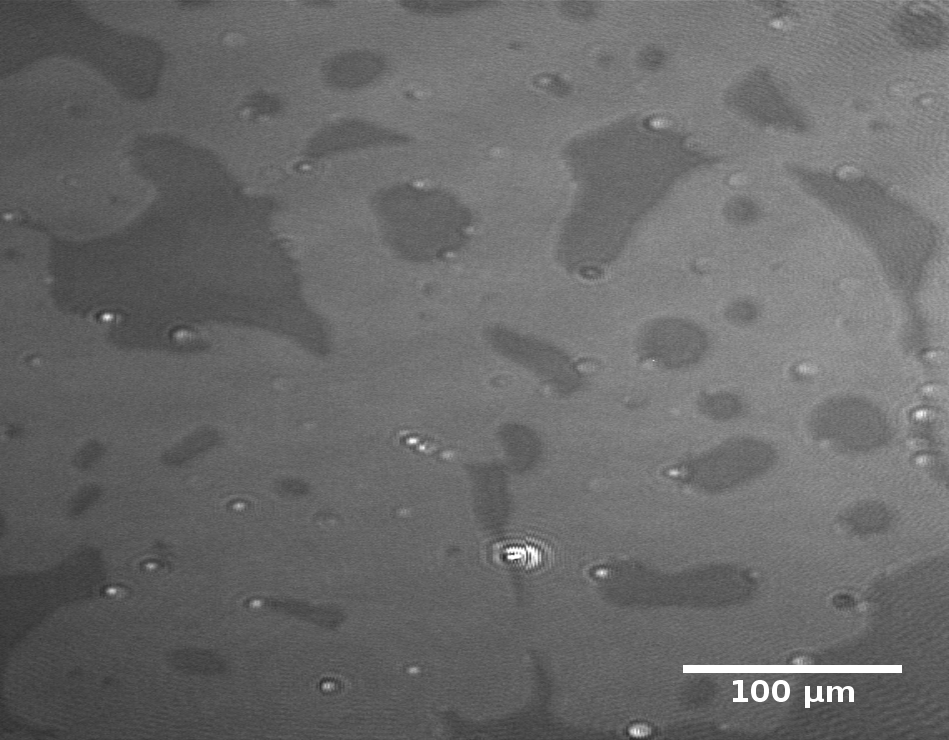}\label{mat40}} \qquad
\subfloat[]{\includegraphics[scale=0.5]{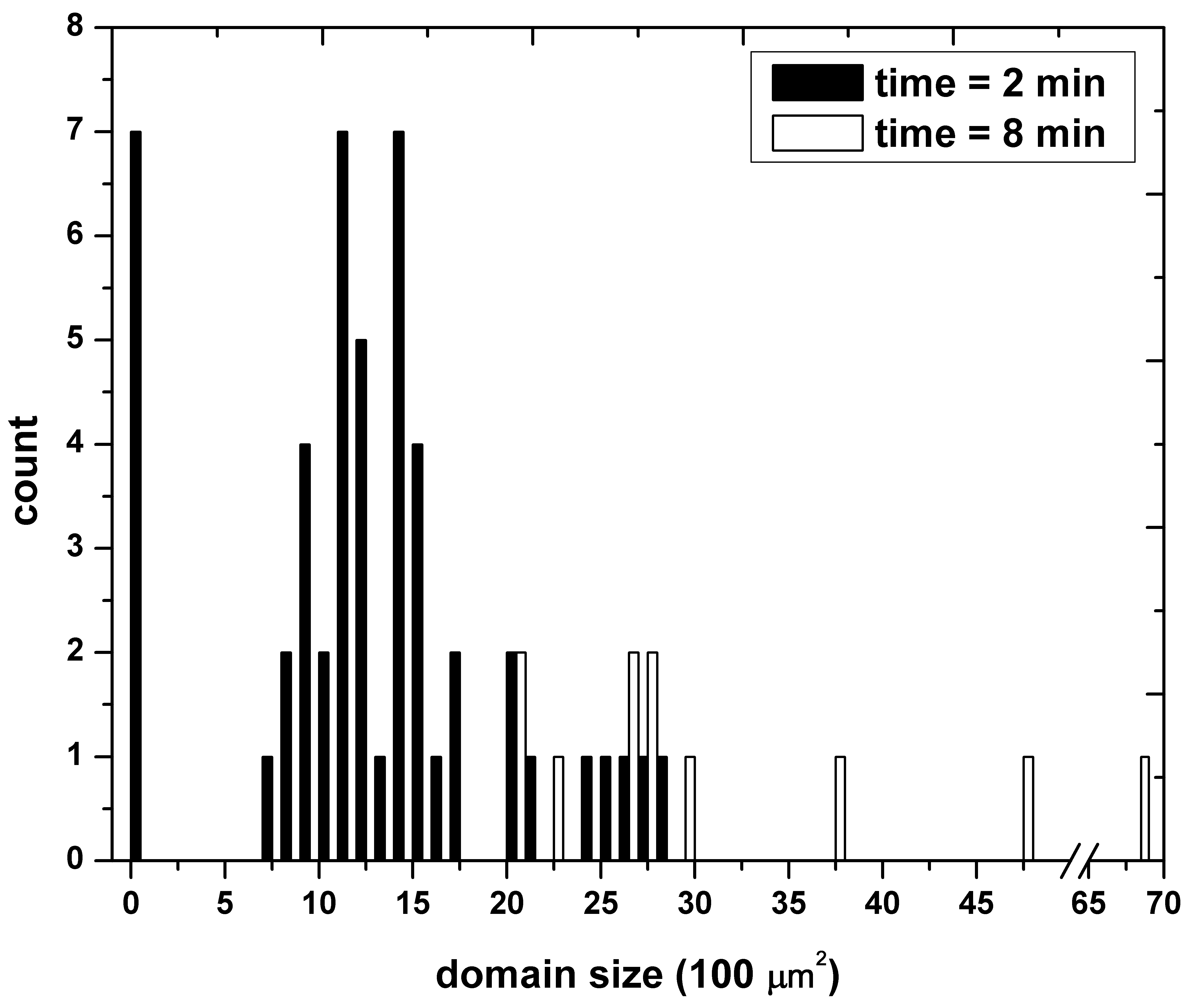}\label{mahist}}
\caption{(a) Area fraction vs time ($A_n - t$) curves of Myristic acid (C14) monolayer at different surface pressures ($\pi$), 25$^{\circ}$C and on pure aqueous subphase (ambient conditions), with corresponding exponential fits (lines, Eqn (2)). See text for details. ((b) and (c)) Brewster Angle Microscope (BAM) images of C14 at $\pi =$ 20 mN/m, 25$^{\circ}$C and (d) distribution of 2D domains for different collapse times.\label{ma}}
\end{figure}

\autoref{ma} \subref{maat} shows area fraction vs time $A_{n} - t$ curves (symbols) for C14 monolayers at different $\pi$-values on a pure aqueous subphase at 25$^\circ$C, as extracted from isotherms. The curves at a particular $\pi$ was obtained by maintaining the monolayer at that $\pi$, measuring the monolayer area as a function of time (t) and normalizing the area values with the initial area. All the curves are fit very well by single exponential decay functions (line) showing that C14 undergoes DD dynamics and can be explained by a simple, semi-empirical model of the desorption mechanism, which assumes that desorption rate depends on the total number of surfactant molecules at the interface.

Let $N$ be the total number of surfactant molecules in a monolayer of area $A$ at any instant $t$. $\pi$ is a function of the concentration of molecules at the interface ($\rho =\frac{N}{A}$). Hence, for a fixed $\pi$, $\rho$ is a constant and $N = \rho A$. From our assumptions,
\begin{equation}
\left(\frac{\Delta N}{\Delta t}\right)_{\pi} = -kN
\end{equation}
leading to
\begin{equation}
A_{n}(t)= A_{i}e^{-kt}
\end{equation}
where $k$ is a decay constant (its reciprocal is time constant $\tau$) which depends on $\rho$ and temperature and $A_{i}$ is the initial monolayer area fraction ($\sim 1.00$). This model is consistent with the fact that desorption occurs at any nonzero pressure.

\autoref{ma} \subref{mat1}-\subref{mat40} show the BAM images of C14 monolayers at $\pi =$ 20 mN/m, 25$^\circ$C  at different relaxation times. At high $\pi$ different 2D domains of C14 are seen to coalesce but no multilayers are formed as is clear from the constant contrast of the BAM images. This indicates dissolution of amphiphiles into the subphase and hence confirms a DD mechanism, consistent with the shorter chain length of C14. The contrast with this model of continous loss of monolayer at macroscopic scale, i.e., from $A_n-t$ curves is brought out clearly in \autoref{ma} \subref{mahist} which demonstrates a simultaneous coalescence of 2D domains and formation of bigger 2D structures in the course of C14 destabilization where the sizes of 2D domains are obtained using ImageJ\cite{schindelin2012} software. From t $\sim$ 2 min to t $\sim$ 8 min, the number of smaller domains decreases and that of bigger domains increases continuously till, after 10 mins, the entire field of view is covered uniformly.

\section{Nucleation Dynamics}

\begin{figure}
\centering
\subfloat[]{\includegraphics[scale=0.5]{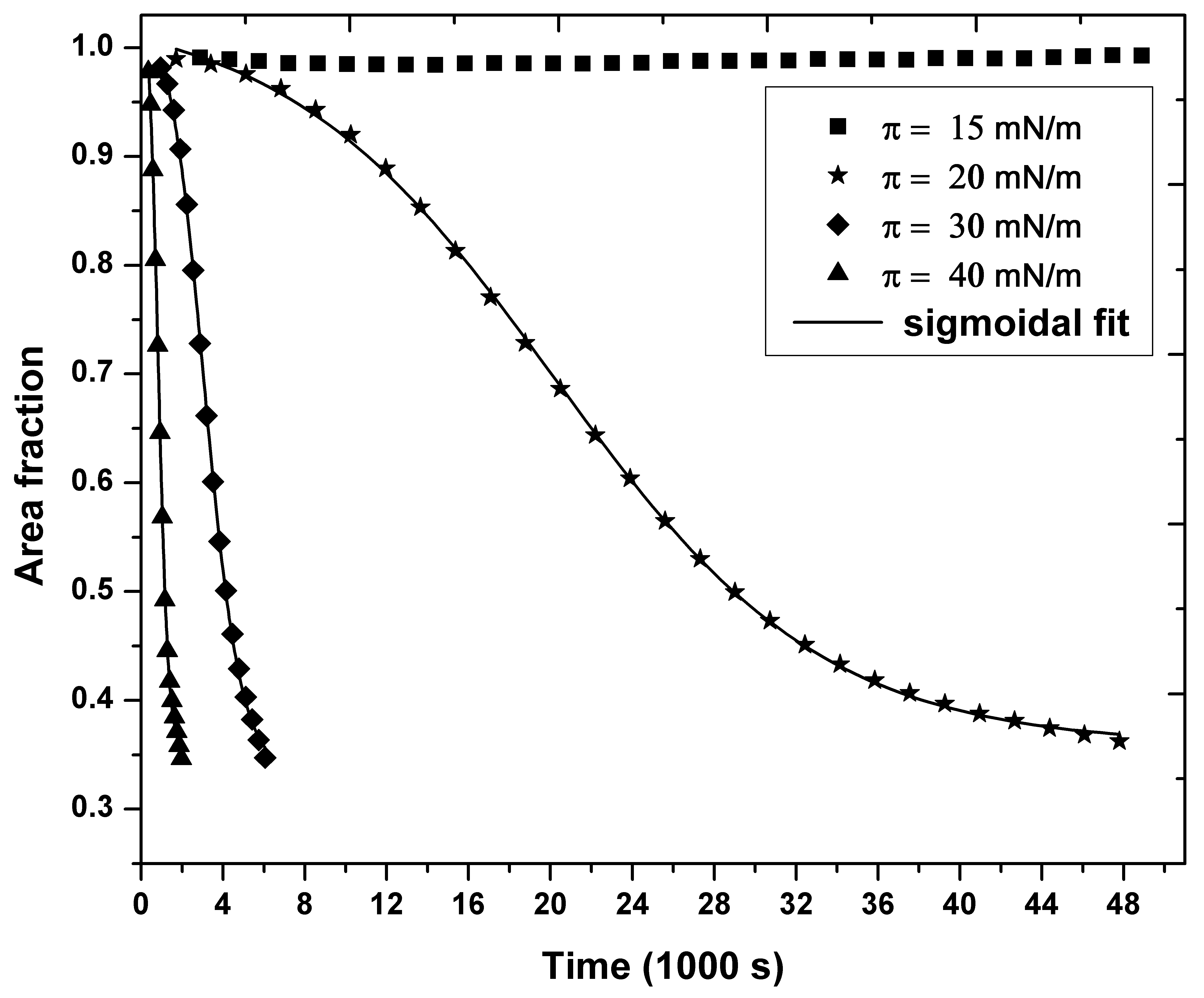}\label{aaat}}\qquad
\subfloat[time $\sim$ 3 min]{\includegraphics[width = 1.9in]{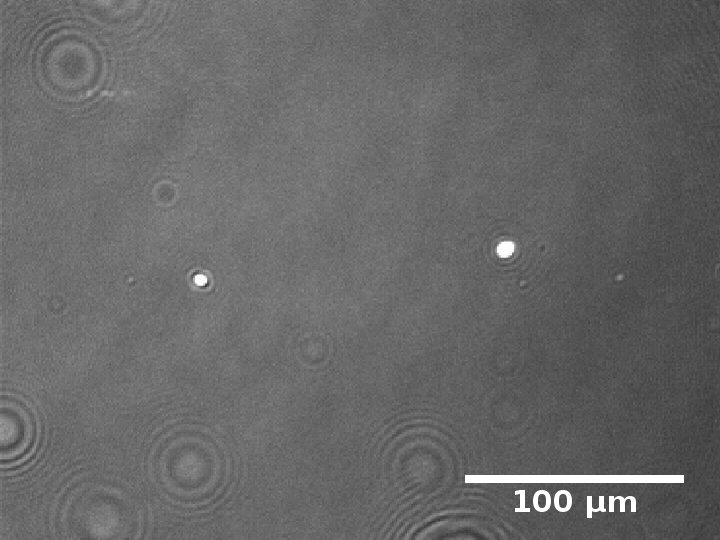}\label{aat3}} \\
\subfloat[time $\sim$ 95 min]{\includegraphics[width = 1.9in]{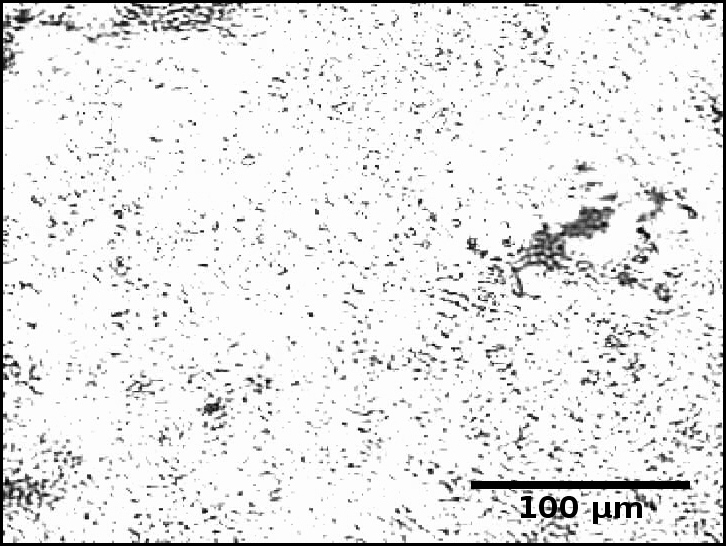}\label{aat95}} \qquad
\subfloat[]{\includegraphics[scale=0.5]{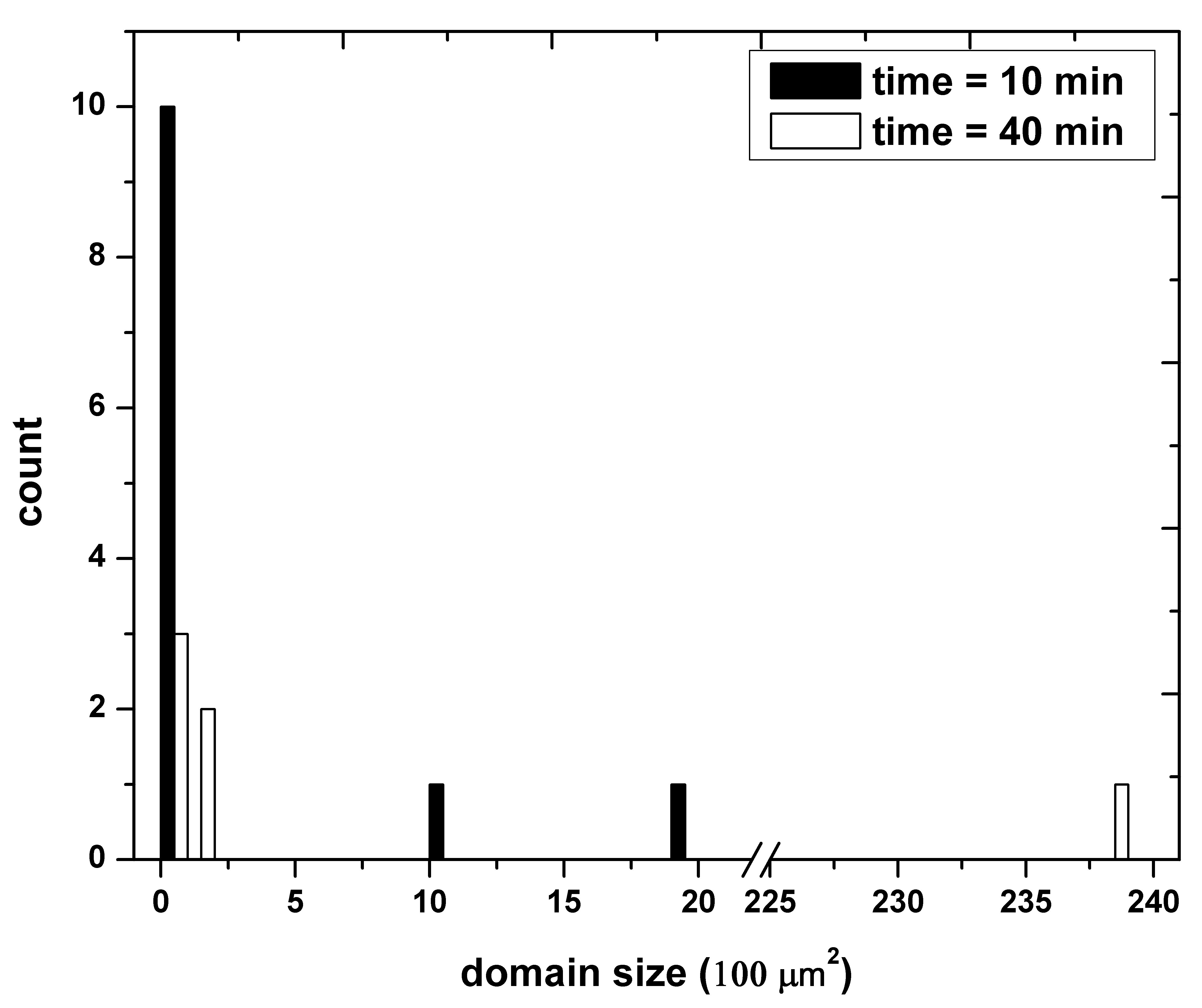}\label{aahist}}
\caption{(a) $A_n - t$ curves of Arachidic acid (C20) monolayers at different surface pressures ($\pi$) and ambient conditions, with sigmoidal fits (lines, Eqn (3)). See text for details. ((b) and (c)) Brewster Angle Microscope (BAM) images of C20 at $\pi =$ 30 mN/m, 25$^{\circ}$C and (d) distribution of 3D domains for different collapse times.\label{aa}}
\end{figure}

$A_{n}- t$ (symbols) curves of C20 monolayers at different values of $\pi$ in \autoref{aa} \subref{aaat} at 25$^\circ$C shows that the curves are sigmoidal indicating ND dynamics for these molecules, as expected from their chain lengths. The monolayer instability increases with $\pi$ and at $\pi \leq$ 15 mN/m the $A_{n}- t$ curves are horizontal indicating a stable monolayer (constant for $\geq$ 13 hrs). The behaviour of C18 is very similar and hence is not shown.

We can model the nucleation mechanism as a self-limiting process. Let $N_{2D}$ and $N_{3D}$ be the total number of surfactant molecules in the 2D monolayer and in the 3D phase respectively at any instant $t$ and $N_{0}$ ($=N_{2D}+N_{3D}$) is the total number of amphiphiles spread initially. Then for growth of the 3D phase
 $$\left(\frac{\Delta N_{3D}}{\Delta t}\right) \propto N_{3D}\left(1-\frac{N_{3D}}{N_{0}}\right)$$ where the term linear in $N_{3D}$ stands for the unimpeded 3D phase growth while the negative quadratic term in $N_{3D}$ models the self-limiting due to depletion in $N_{2D}$. In terms of $N_{2D}(=N_{0}-N_{3D})$, we have for decay of 2D phase
  $$\left(\frac{\Delta N_{2D}}{\Delta t}\right) \propto - N_{2D}\left(1-\frac{N_{2D}}{N_{0}}\right)$$
For fixed $\pi$, $\rho$ is a constant and $N_{2D_{(e)}}=\rho A$ where $N_{2D_{(e)}}$ includes the monolayer and also the base layer of the multilayers in contact with the water subphase. We assume that growth of multilayers stops only when the monolayer is totally depleted. Thus $N_{2D_{(e)}}$ can be considered to be $N_{2D}$\cite{kundu2006}. Then we have
\begin{align*}
 \left(\frac{\Delta A}{\Delta t}\right)_{\pi} &\propto - A\left(1-\frac{A}{A_{0}}\right)
 \end{align*}
or
\begin{equation}
A_{n}(t) = A_{f}+\frac{A_{i}-A_{f}}{1+e^{k(t-t_{0})}}
\end{equation}
where $A_{0}$, $A_{f}$, $k$ and $t_{0}$ are the initial monolayer area, final area fraction ($\sim 0.2$), decay constant (measuring the steepness of the curves) and inflection point (2D-3D coexistence time), respectively. This function has been used to fit the relaxation curves of nucleation in \autoref{aa} \subref{aaat} (line).

 \autoref{aa} \subref{aat3}-\subref{aat95} are the BAM images of C20 monolayers at different times during relaxation at $\pi$ = 30 mN/m and 25$^\circ$C. It is evident that these relaxations, unlike C14, correspond to monolayer to multilayer transformation, i.e. nucleation, starting from multilayer centres shown as bright dots formed randomly over the entire monolayer, which grow and coalesce along with growth of new multilayer centres (\autoref{aa} \subref{aat3}), till the entire monolayer is transformed into multilayers (\autoref{aa} \subref{aat95}) confirming this to be a ND process. However, the time for monolayer to multilayer transformation is different for C20 and C18 monolayers at the same $\pi$. Comparison of BAM and isotherm data indicates a clear correlation between nucleation and sigmoidal decay but it is to be noted that these sigmoidal profiles are observed only when the data are taken over a significantly long time and hence they have not been observed before, to the best of our knowledge. \autoref{aa} \subref{aahist} shows the distribution of 3D phases over C20 monolayer as it relaxes at $\pi$ = 30 mN/m and 25$^\circ$C. From t = 10 min to 40 min, the area of the largest 3D domain is found to grow from 2000 $\mu m^{2}$ (\autoref{aa} \subref{aahist}) to 23800 $\mu m^{2}$ (\autoref{aa} \subref{aahist}), coexisting with a few smaller domains.

BAM studies show that in-plane coalescence is the common and crucial process in both desorption and nucleation dynamics. The importance of 2D coalescence in ND dynamics has been shown through previous BAM studies\cite{vollhardt2006} but, to our knowledge, its presence and importance in DD dynamics has not been discussed before. We suggest that these results point to the importance of lipophilic force in monolayer dynamics on a long time scale.

\section{Control Parameters}
\subsection{Surface Pressure}

\begin{figure}
\centering
 \subfloat[]{\includegraphics[scale=0.7]{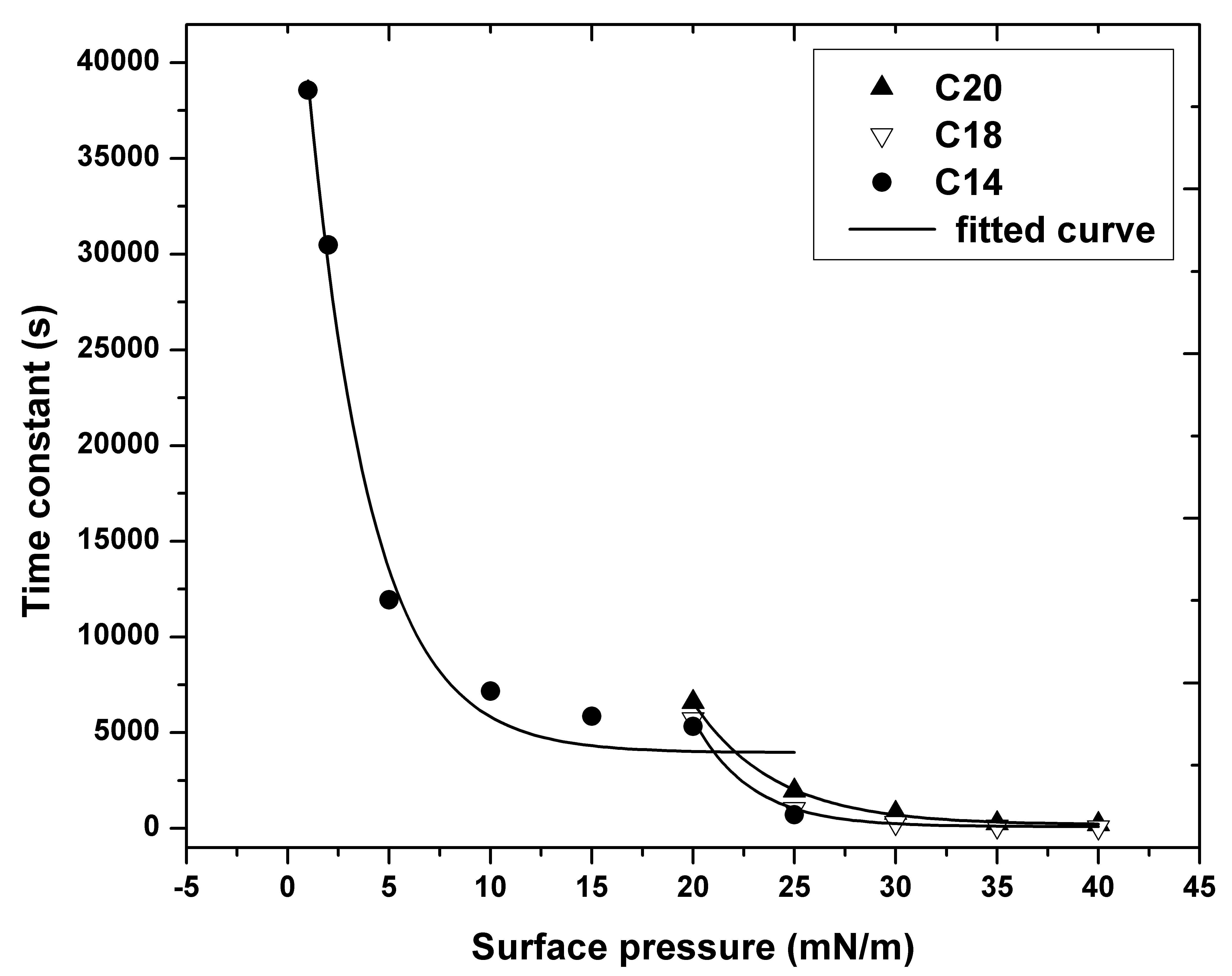}\label{pit}}\quad
 \subfloat[]{\includegraphics[scale=0.7]{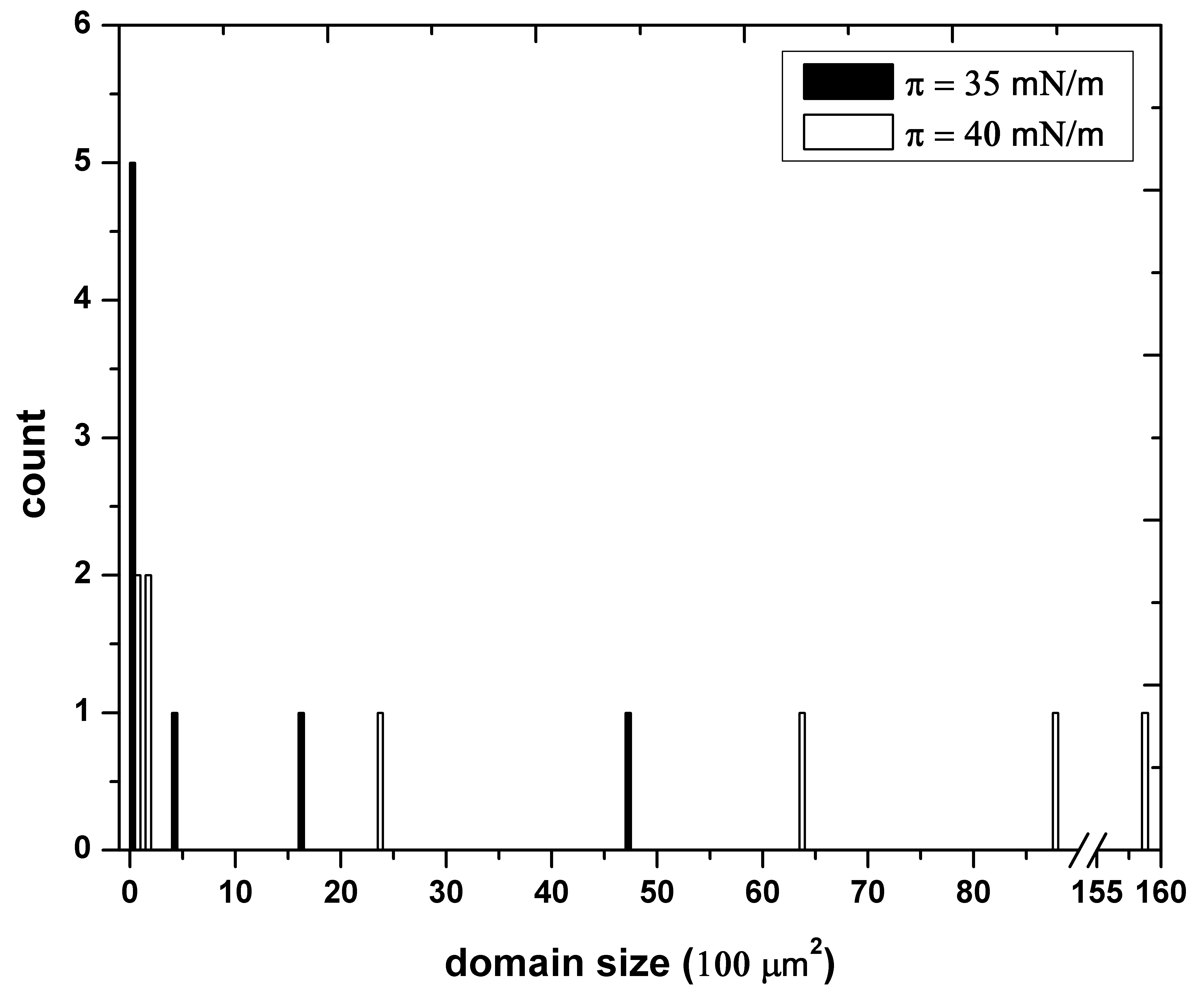}\label{pihist}}
 \caption{(a) Time constants ($\tau$'s), obtained by fitting $A_n - t$ curves of C14 with eqn (2) and those of C18 and C20 with eqn (3), at different surface pressures and ambient conditions. (b) Distribution of 3D domains from BAM images of C20 at 35 mN/m and 40 mN/m and ambient conditions, after 15 mins of relaxation.\label{pi}}
 \end{figure}

The time constants ($\tau = k^{-1}$) of C14, C18 and C20 monolayers as functions of surface pressure at 25$^\circ$C are shown in \autoref{pi} \subref{pit}. $\tau$ may be treated as a measure of stability of the monolayer against hydrophilic and lipophilic forces in desorption and nucleation, respectively, and is obtained from the fits of equations (2) and (3) to the respective $A_{n}- t$ curves. $\tau$ of C14 decreases exponentially with surface pressure. This enhanced desorption for larger 2D clusters suggests a correlated diffusion or superdiffusion, possibly due to the interaction between the `sheet' of dipoles (anions) from undissociated (dissociated) headgroups and water \cite{duncan1991}.

Vysotsky et al observed in their series of papers the spontaneous clustering of aliphatic amides for alkyl chain length higher than a threshold value, forming dimers and tetramers\cite{vysotsky2011,vysotsky2012,vysotsky2013}. Goto et al showed that during compression the hydrophilic groups are protruded in a new geometric configuration to form trilayer (or multilayer) structure\cite{goto2013}. Our results for the longer chain acids are consistent with their studies. The stability of C20 and C18 decreases exponentially with surface pressure probably because molecules can then come closer to form dimers, thereby turning hydrophobic and diffusing upwards to form multilayers.

\autoref{pi} \subref{pihist} shows the distribution of domain areas from BAM images of C20 monolayers at 35 mN/m and 40 mN/m, respectively, after 15 min of relaxation. The maximum domain area grows from $\sim$ 4800 $\mu m^{2}$ to $\sim$ 16000 $\mu m^{2}$, clearly indicating 2D coalescence besides 3D nucleation.

\subsection{Temperature}

\begin{figure}
\centering
 \subfloat[  ]{ \includegraphics[scale=0.7]{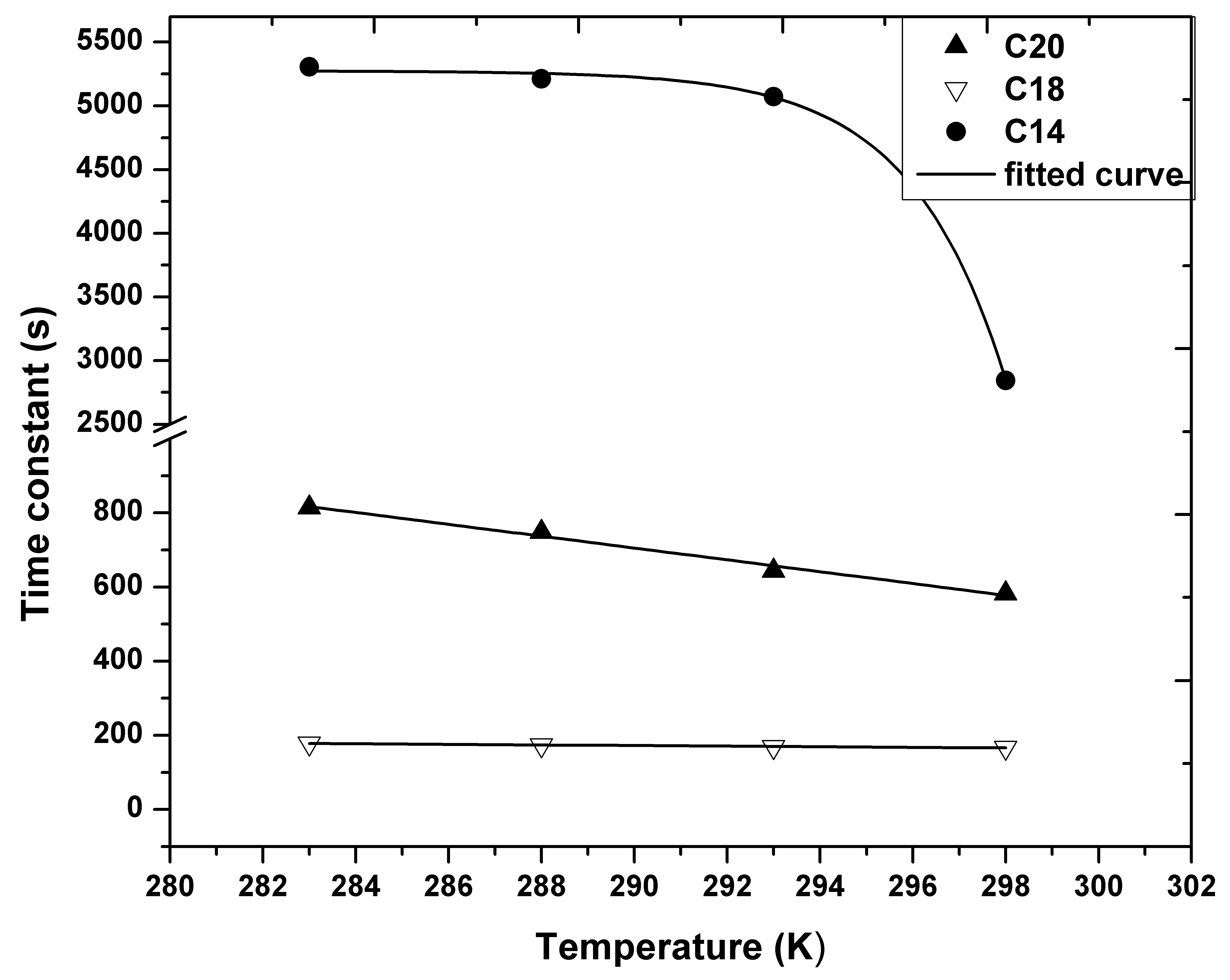}\label{tempt}}\quad
 \subfloat[]{\includegraphics[scale=0.7]{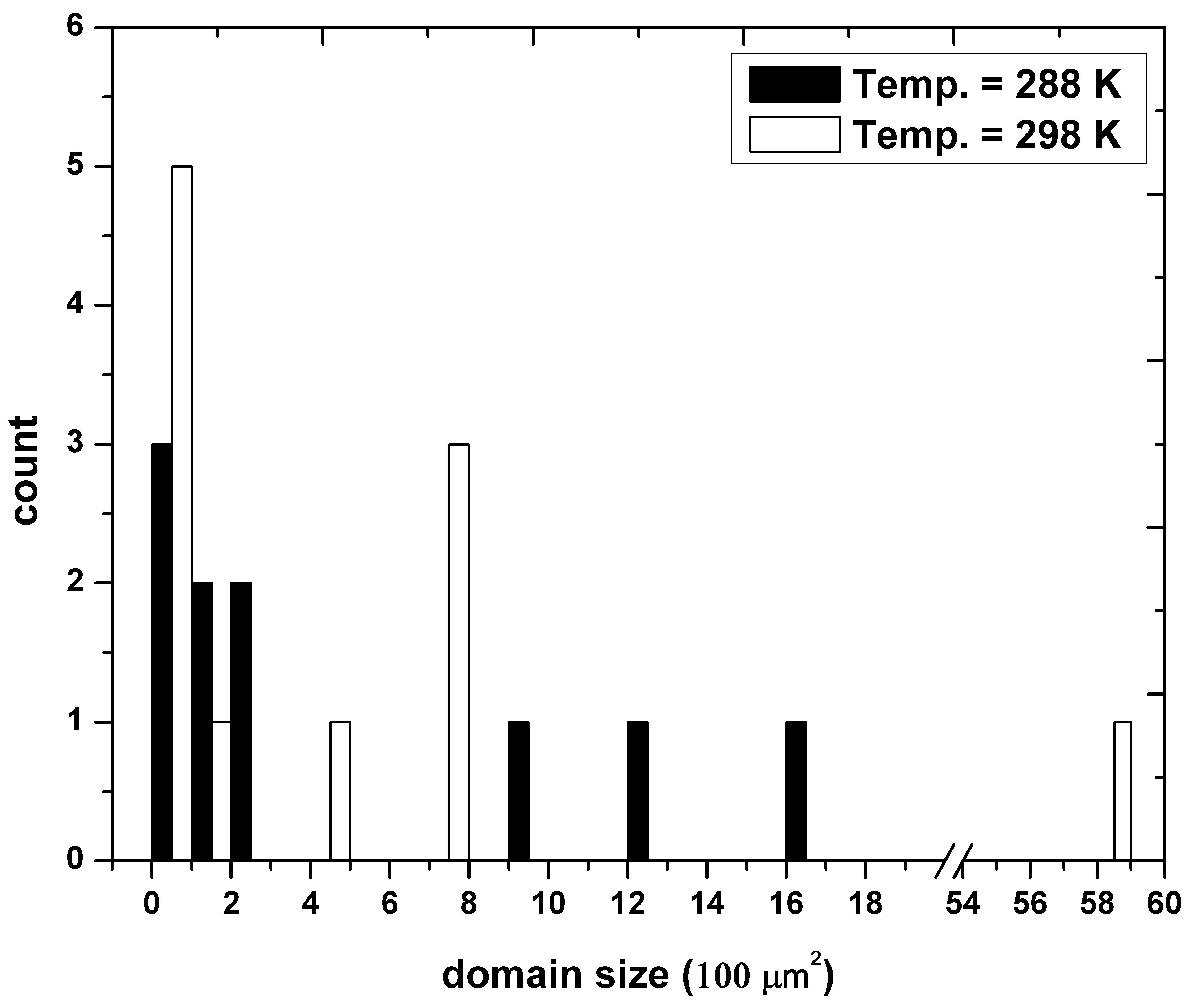}\label{temphist}}
 \caption{(a) $\tau$'s, obtained by fitting $A_n - t$ curves of C14 with eqn (2) and those of C18 and C20 with eqn (3), at different temperatures and 35 mN/m on pure water. (b) Distribution of 3D domains from BAM images of C20 at 288 K and 298 K and above conditions, after 25 mins of relaxation.\label{temp}}
 \end{figure}

The time constants of C14, C18 and C20 monolayers at $\pi =$ 20 mN/m as functions of temperature are shown in \autoref{temp} \subref{tempt}. The stability of C14  remains constant up to 293 K and then falls down rapidly, which indicates overcoming the surface barrier due to tail-water hydrophobic repulsion. This gives a value of this barrier around 25 meV - a remarkably low value, again suggesting a correlation between the headgroups and water that acts against the hydrophobic repulsion.

The stability of C20 and C18 decreases linearly with the temperature but this dependence is weak relative to that of C14 with temperature, as shown in Table I. The weak dependence of C20 monolayer dynamics on temperature can be also seen from BAM studies at 35 mN/m after 25 min of relaxation (\autoref{temp} \subref{temphist}). The number of domains within 0-1000 $\mu m^{2}$ remains very small and the area of the largest domain increases from 1600 $\mu m^{2}$ at 15$^\circ$C to 5850 $\mu m^{2}$  at 25$^\circ$C, showing moderate 2D coalescence.

\section{Cross-over of Dynamics}

\begin{figure}[H]
\centering
\subfloat[]{\includegraphics[scale=0.65]{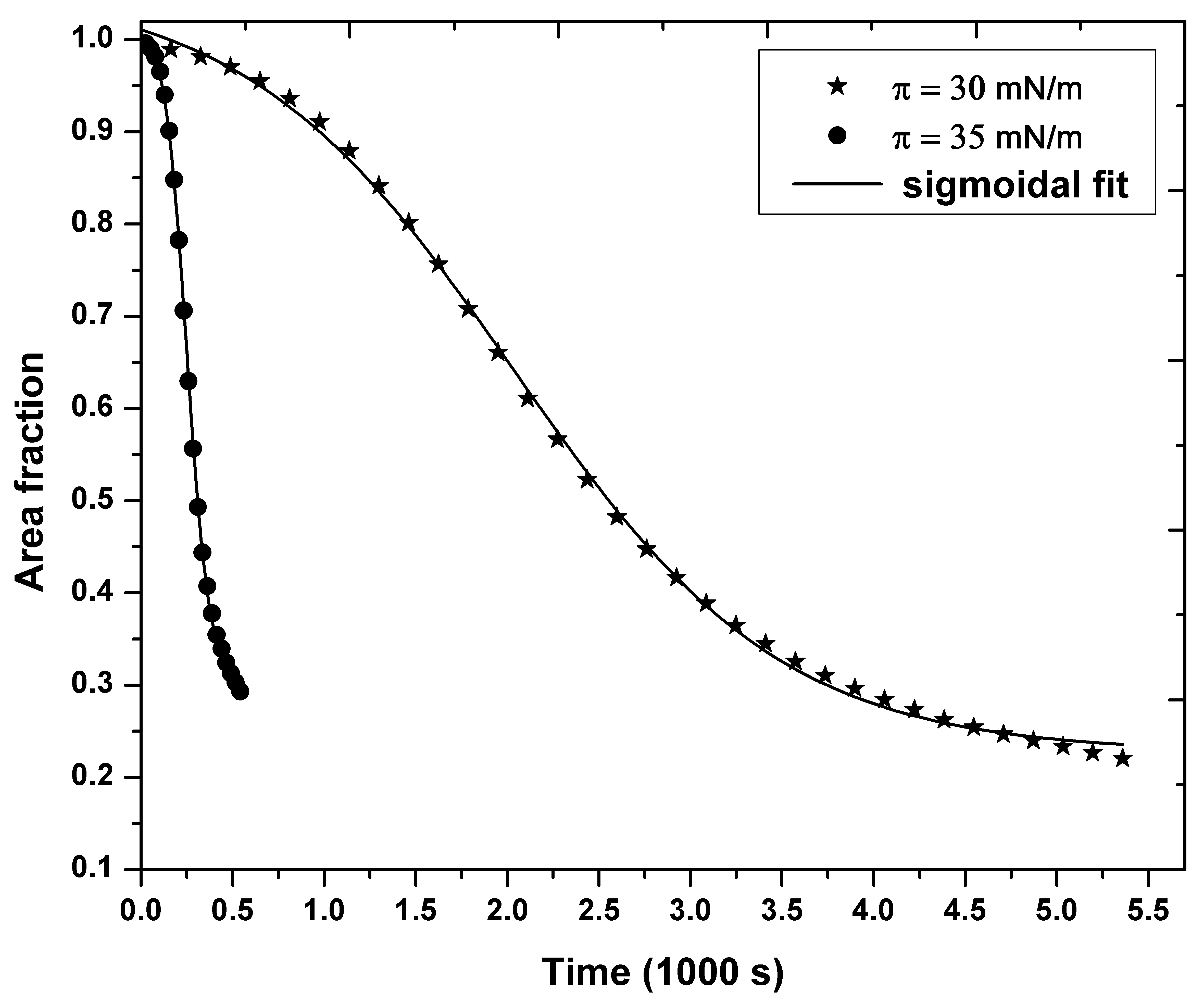}\label{paat1}} \qquad
\subfloat[]{\includegraphics[scale=0.65]{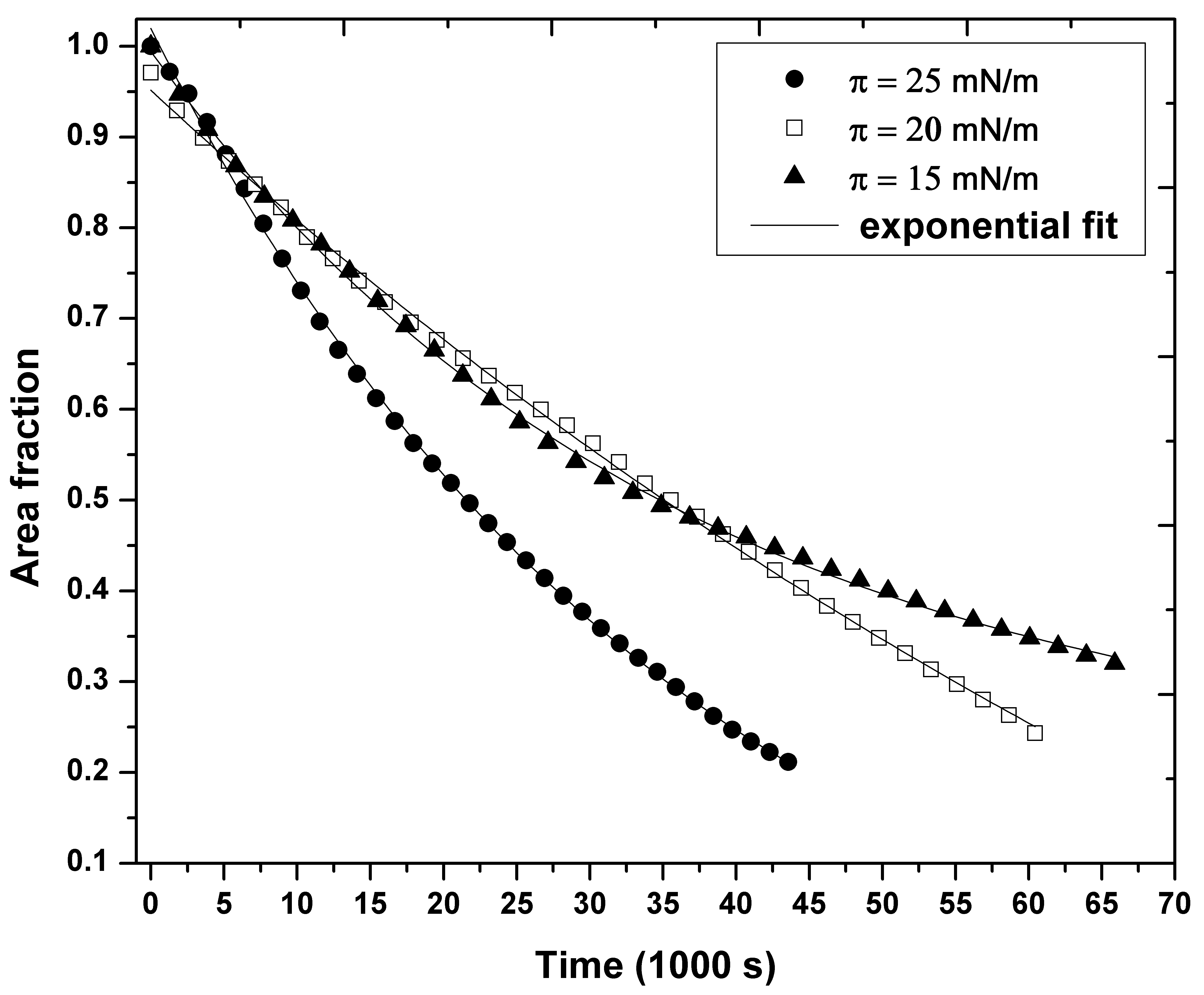}\label{paat2}} \\
 \subfloat[time $\sim$ 90 min]{\includegraphics[width = 1.9in]{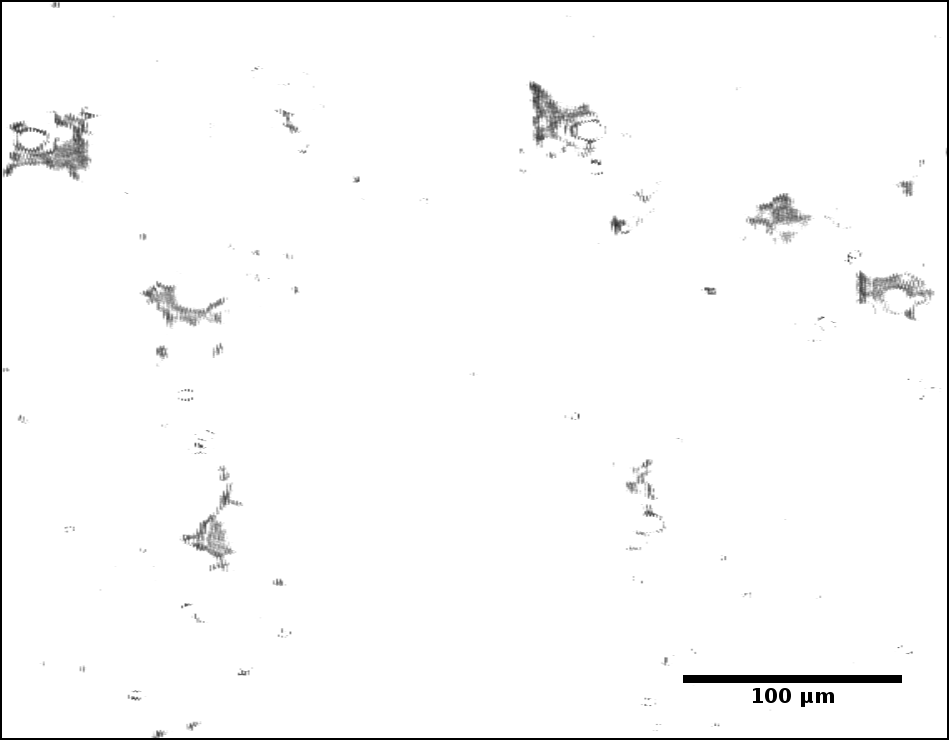}\label{pat90}}\qquad \qquad
 \subfloat[time $\sim$ 10 h]{\includegraphics[width = 1.9in]{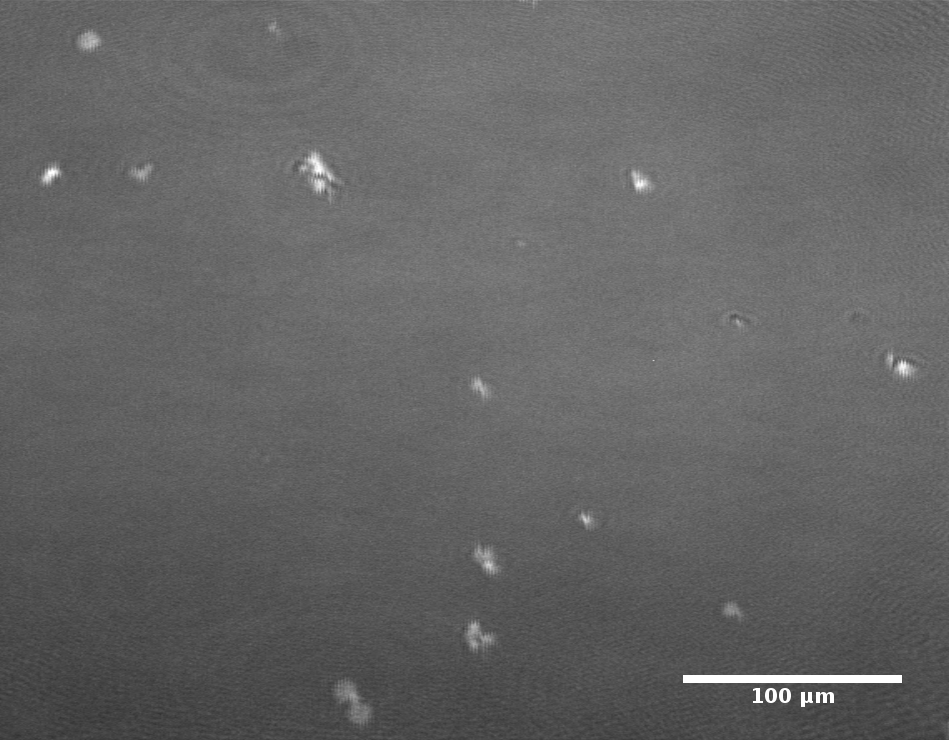}\label{pat10h}}
 \caption{$A_n - t$ curves of  Palmitic acid (C16) monolayers at high and low $\pi$ and ambient conditions, with (a) sigmoidal fits for high pressures and (b) exponential fits for low pressures. BAM images of C16 monolayer at (c) $\pi =$ 30 mN/m and (d) $\pi =$ 1 mN/m for different collapse times. \label{pa}}
 \end{figure}

The relaxation curves of C16 at 25$^\circ$C are sigmoidal at higher surface pressures (\autoref{pa} \subref{paat1}) but exponential at lower pressures (\autoref{pa} \subref{paat2}) showing that C16 monolayer, with tail length between C14 and C18, destabilizes via both ND and DD mechanisms. Above the critical surface (density) of 20 mN/m the enhanced lipophilic attraction makes nucleation dominant leading to sigmoidal shape of the transformation curves. Below it the attraction drops and the dominant mechanism of destabilization starts to be desorption. Around 20 mN/m, the $A_{n}-t$ curves are of a nature intermediate between the processes and the contribution of each is sensitive to changes in $\pi$.

These results are borne out by the BAM images of C16 during constant pressure relaxation at high (30 mN/m, \autoref{pa} \subref{pat90}) and low (1 mN/m, \autoref{pa} \subref{pat10h}) $\pi$. The C16 monolayer collapses completely at $\pi =$ 30 mN/m via 2D to 3D transition after 90 min, while no multilayer is formed even after 10 h at $\pi =$ 1 mN/m and some material disappeared from the interface during the constant pressure relaxation. This is the clearest evidence of the importance of tail length and lipophilic interactions.

\section{Out-of-plane growth in Arachidic Acid Monolayers}
\begin{figure}[H]
\centering
\subfloat[t = 30 min]{\includegraphics[scale=1.5]{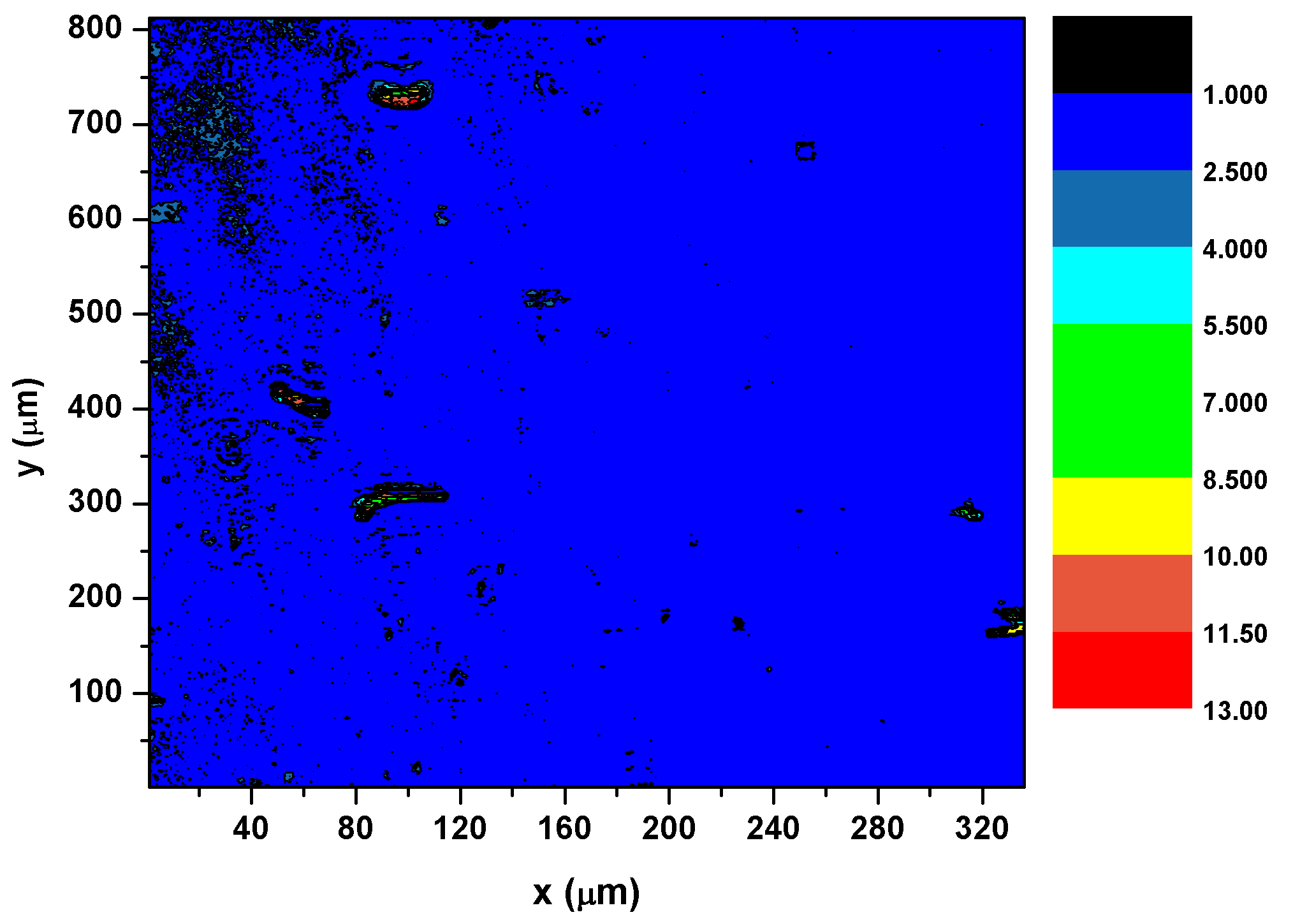}\label{aaT30}}\qquad
\subfloat[t = 30 min]{\includegraphics[scale=1.3]{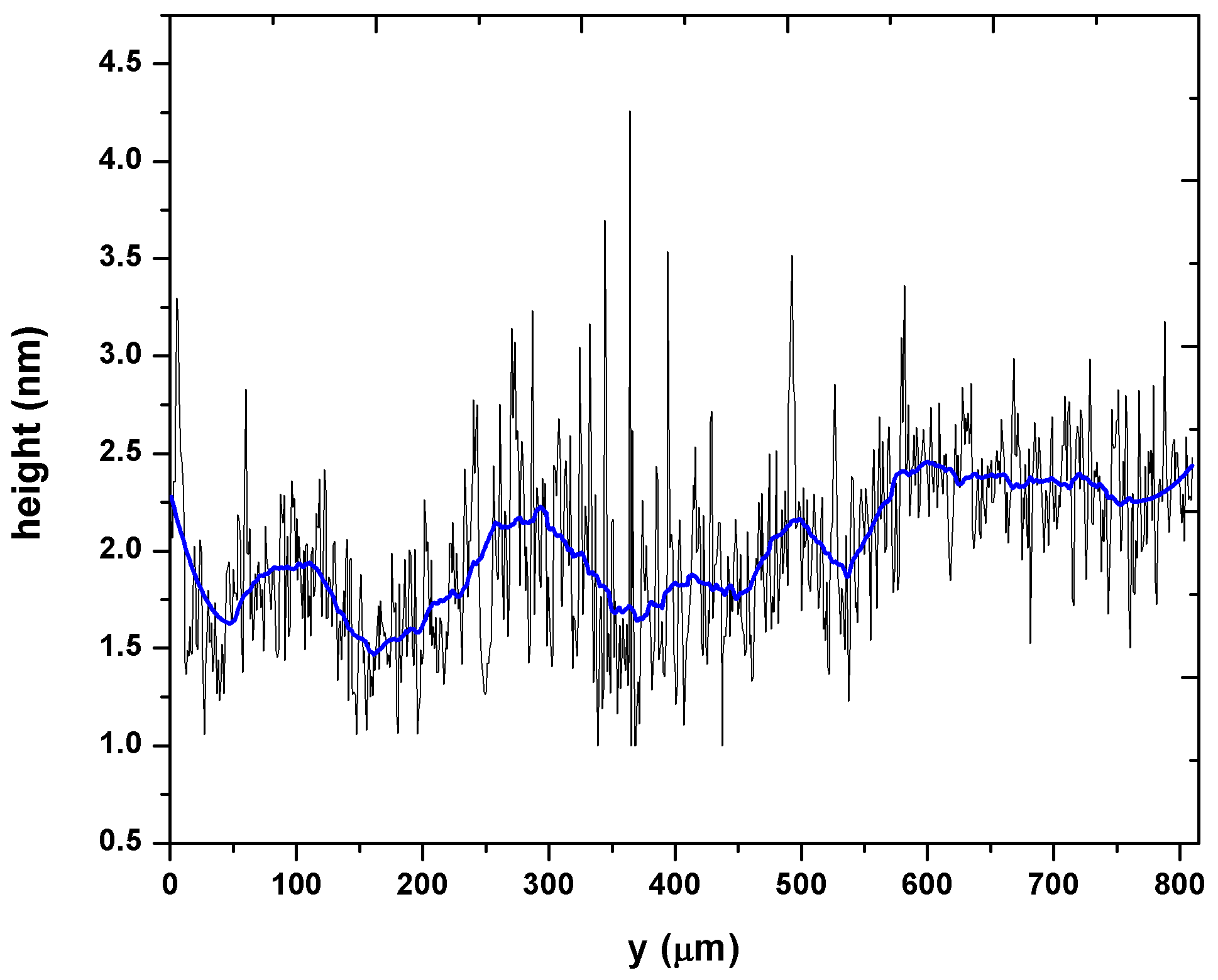}\label{lineT30}}\qquad
\subfloat[t = 60 min]{\includegraphics[scale=1.5]{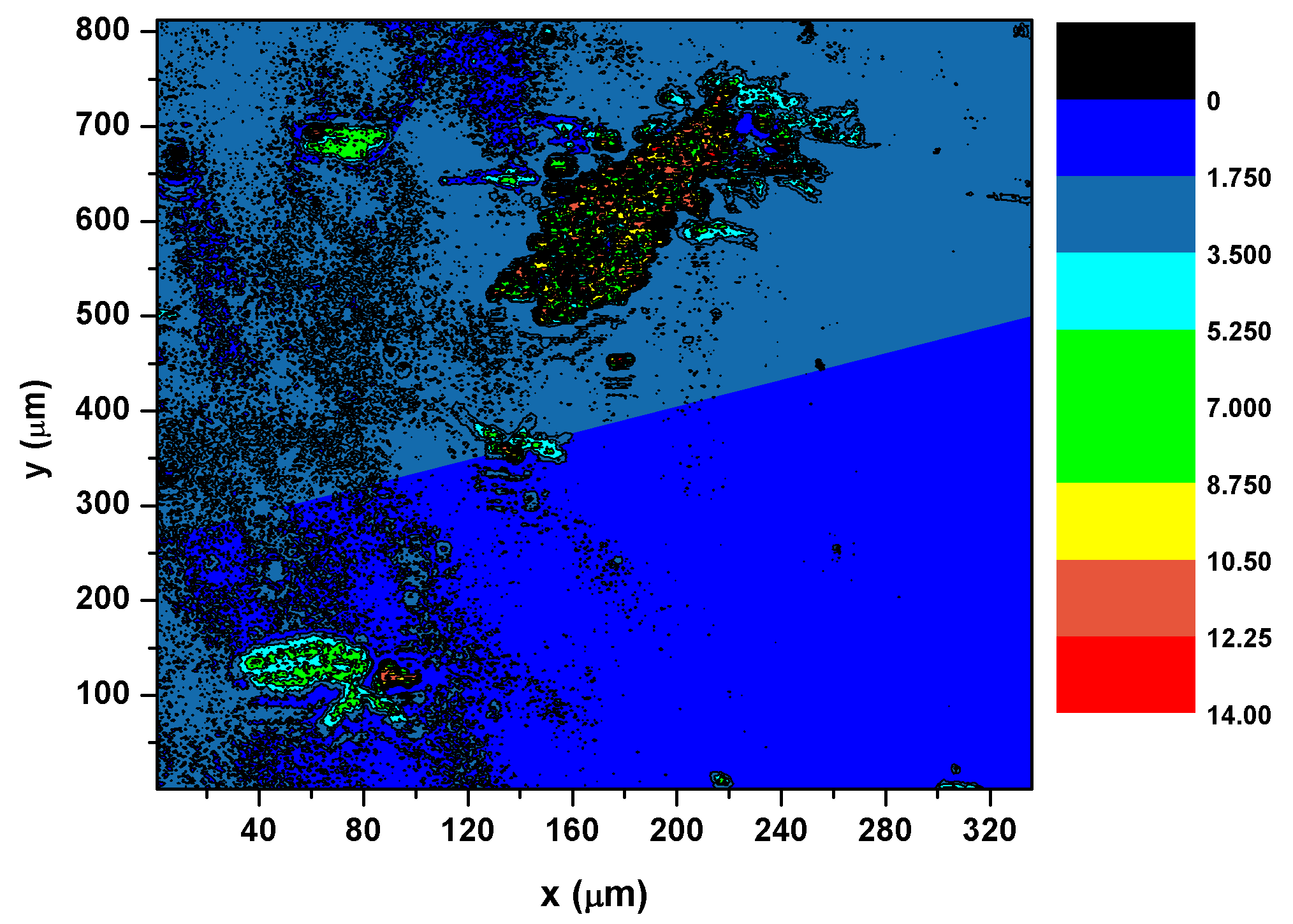}\label{aaT60}}\qquad
\subfloat[t = 60 min]{\includegraphics[scale=1.3]{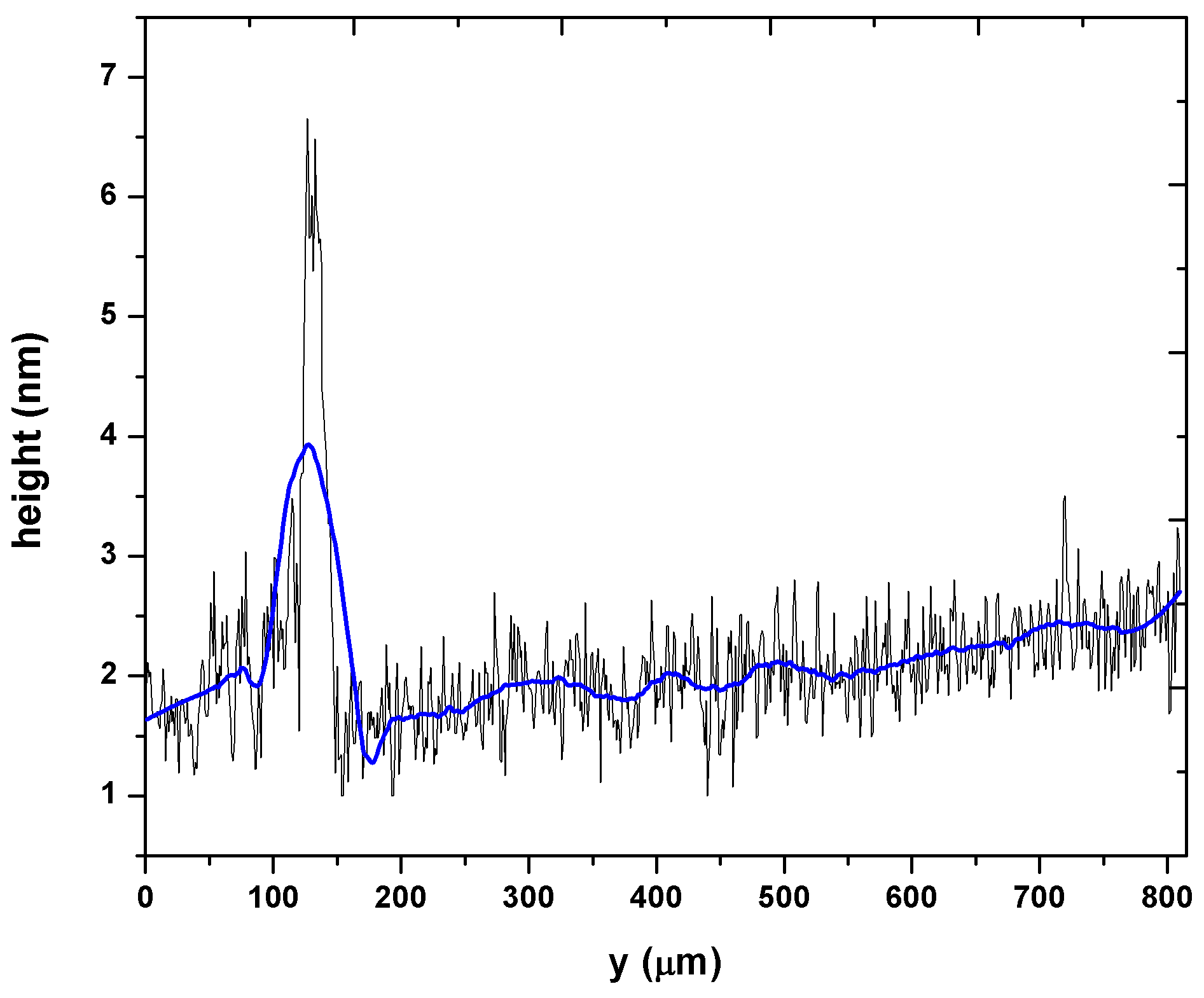}\label{lineT60}}\qquad
\subfloat[t = 90 min]{\includegraphics[scale=1.5]{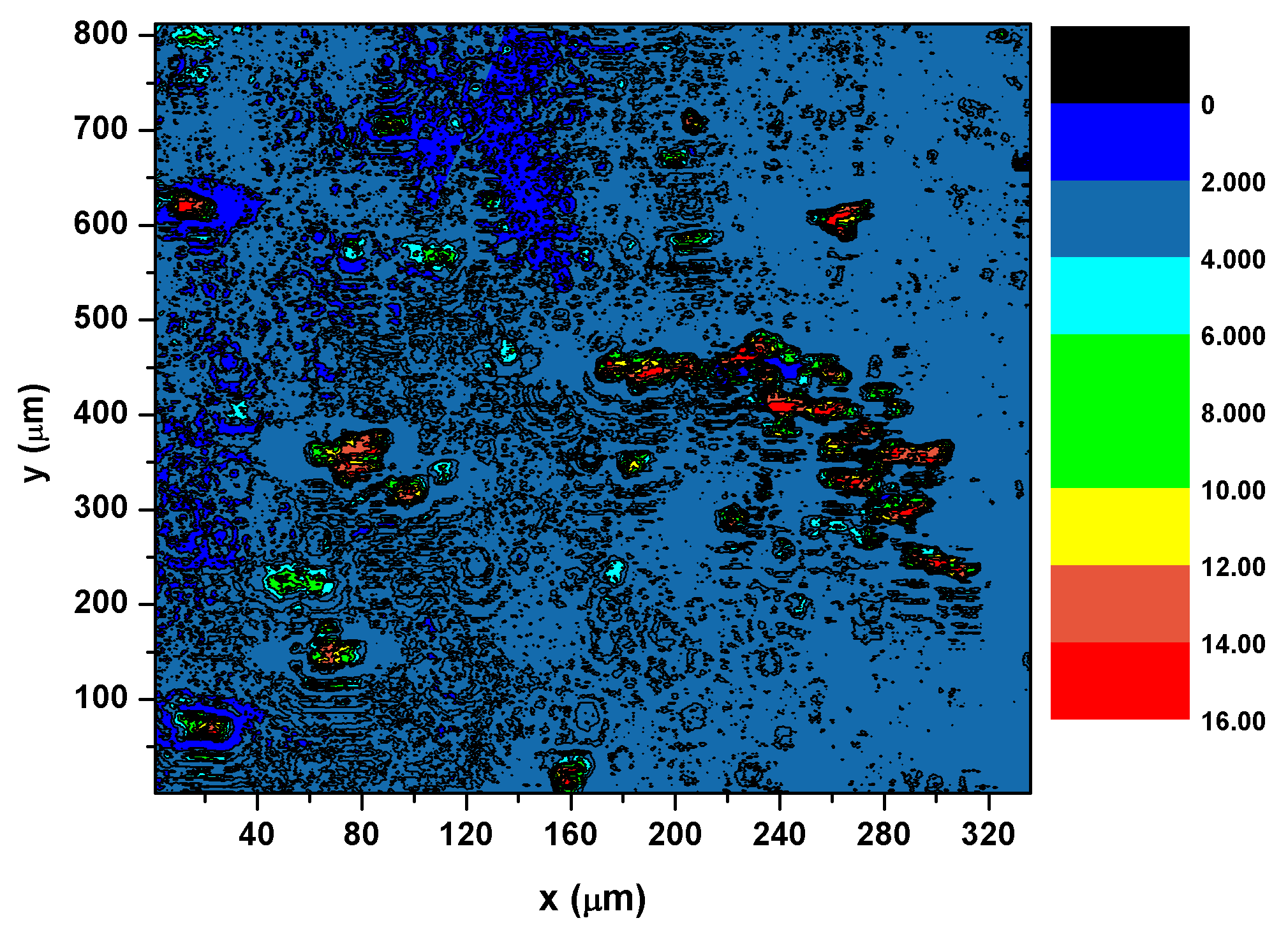}\label{aaT90}}\qquad
\subfloat[t = 90 min]{\includegraphics[scale=1.3]{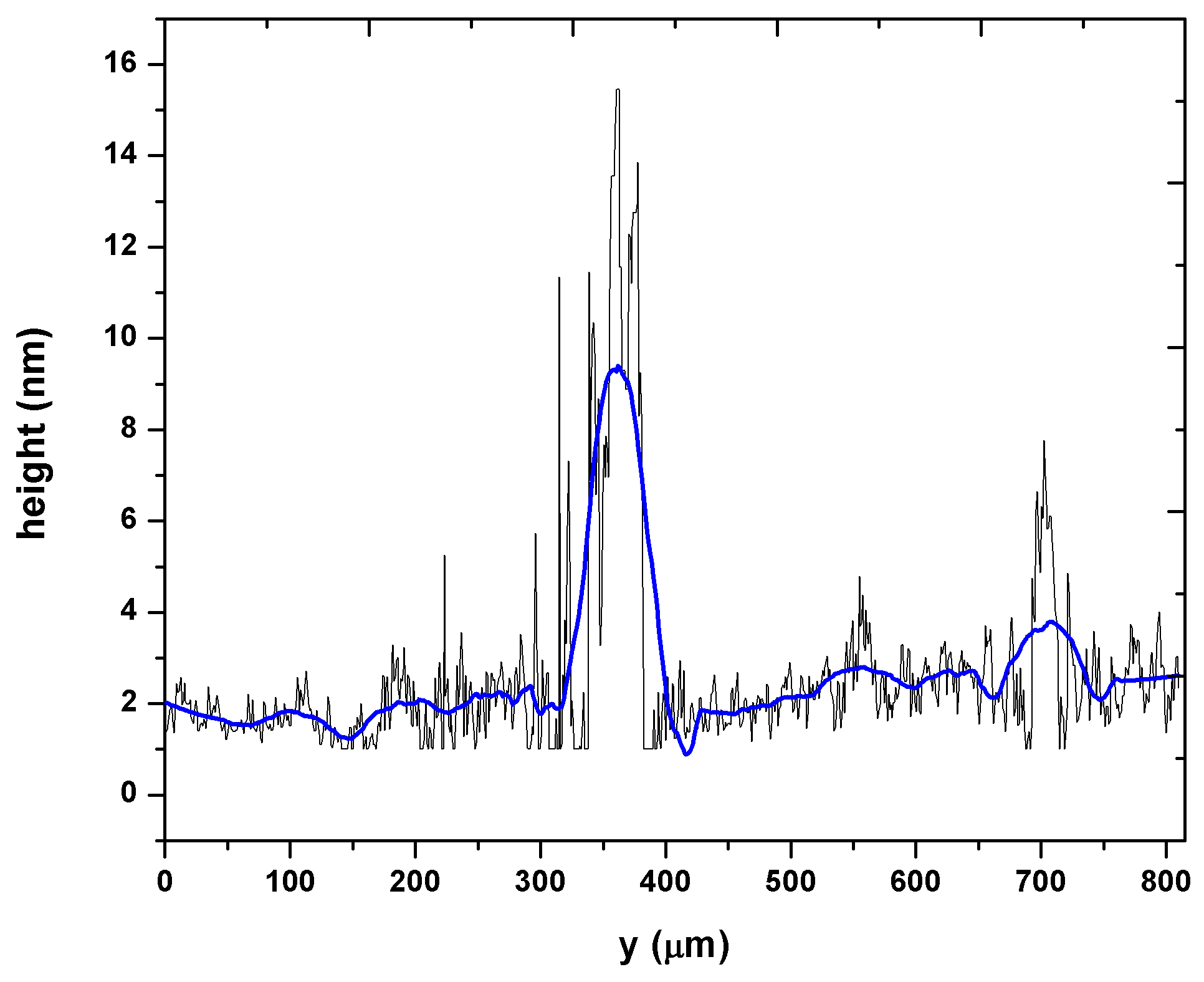}\label{lineT90}}
\end{figure}

\begin{figure}[H]
\centering
\subfloat[t = 180 min]{\includegraphics[scale=1.5]{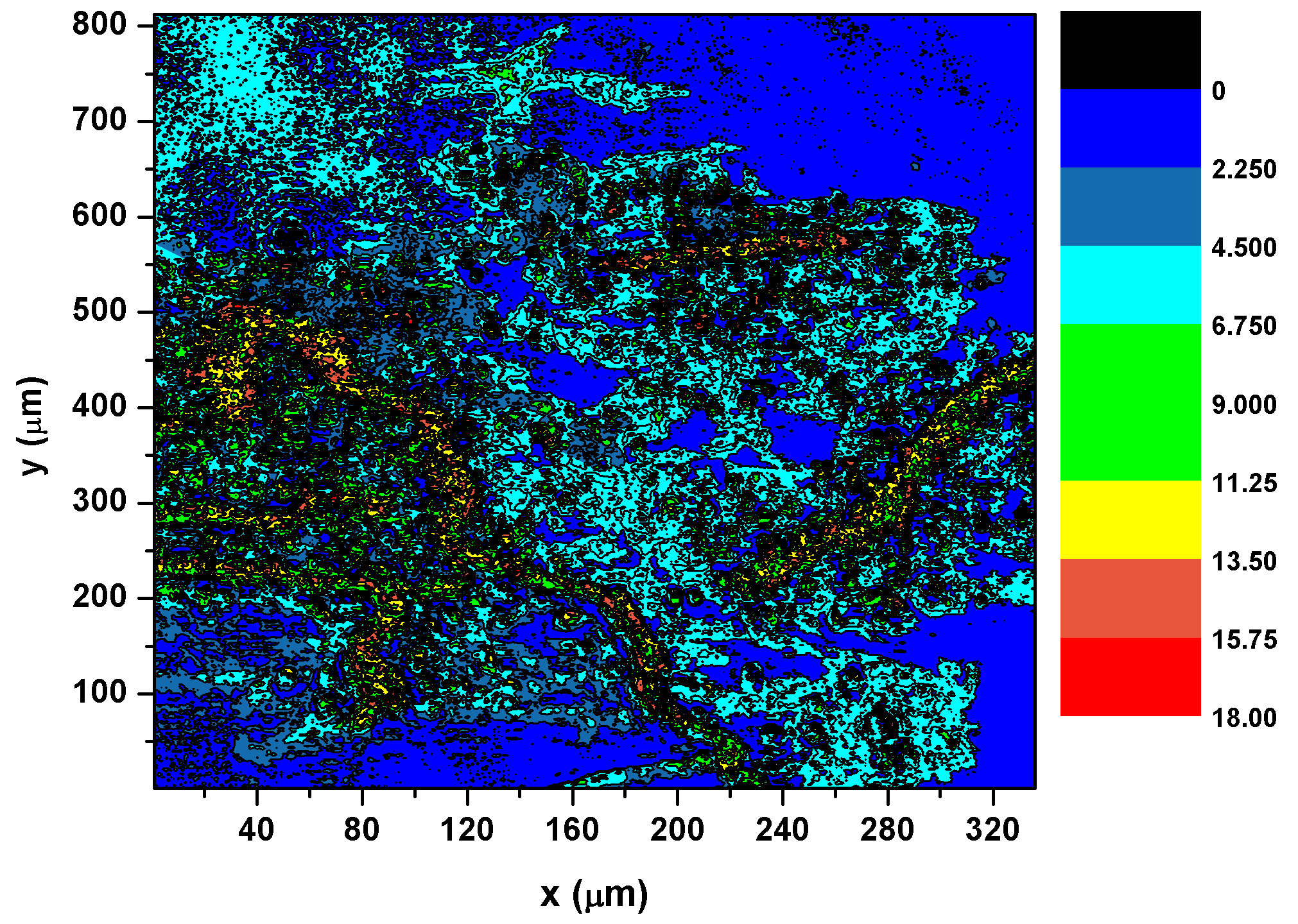}\label{aaT180}}
\subfloat[t = 180 min]{\includegraphics[scale=1.3]{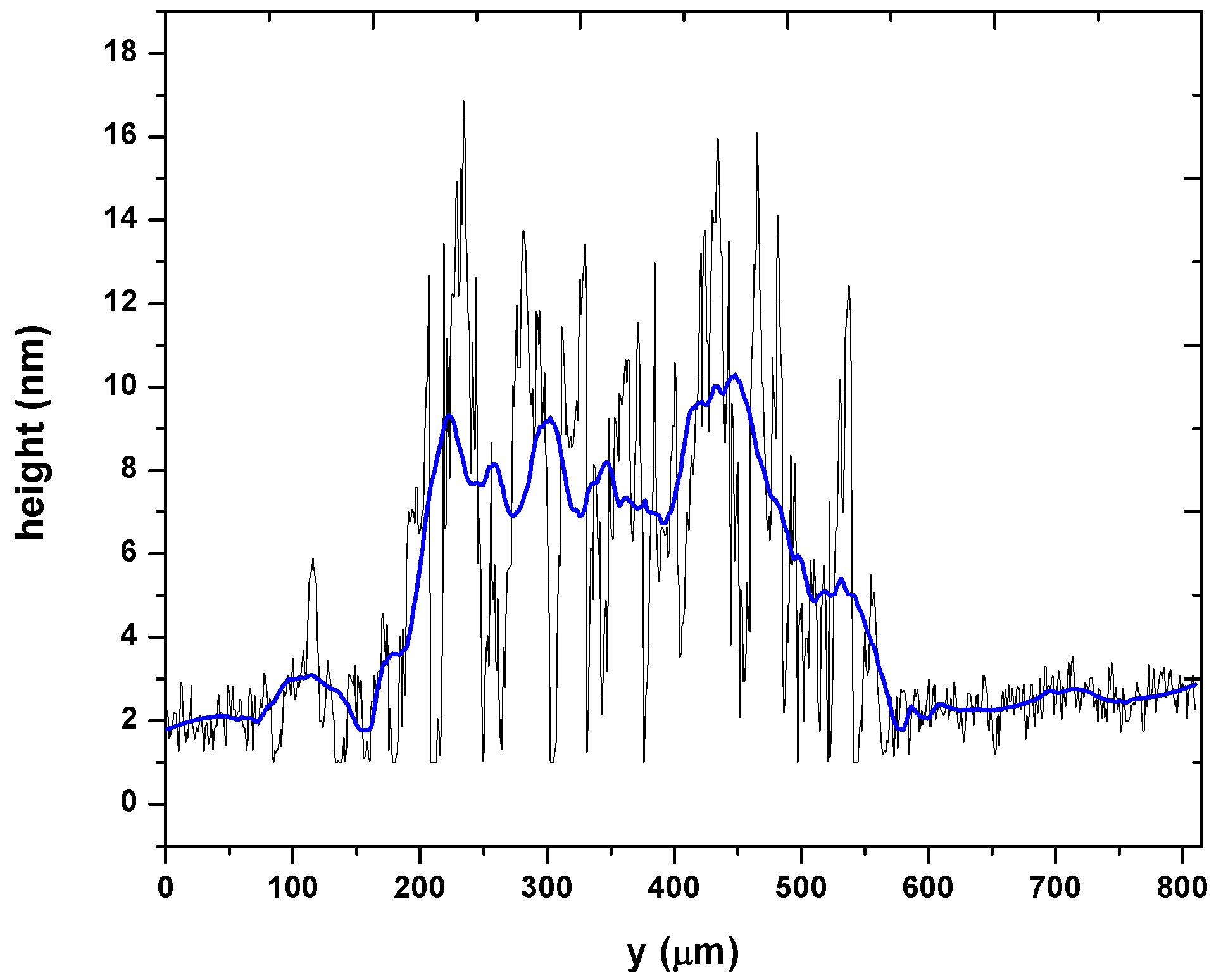}\label{lineT180}}\qquad
\subfloat[t = 240 min]{\includegraphics[scale=1.5]{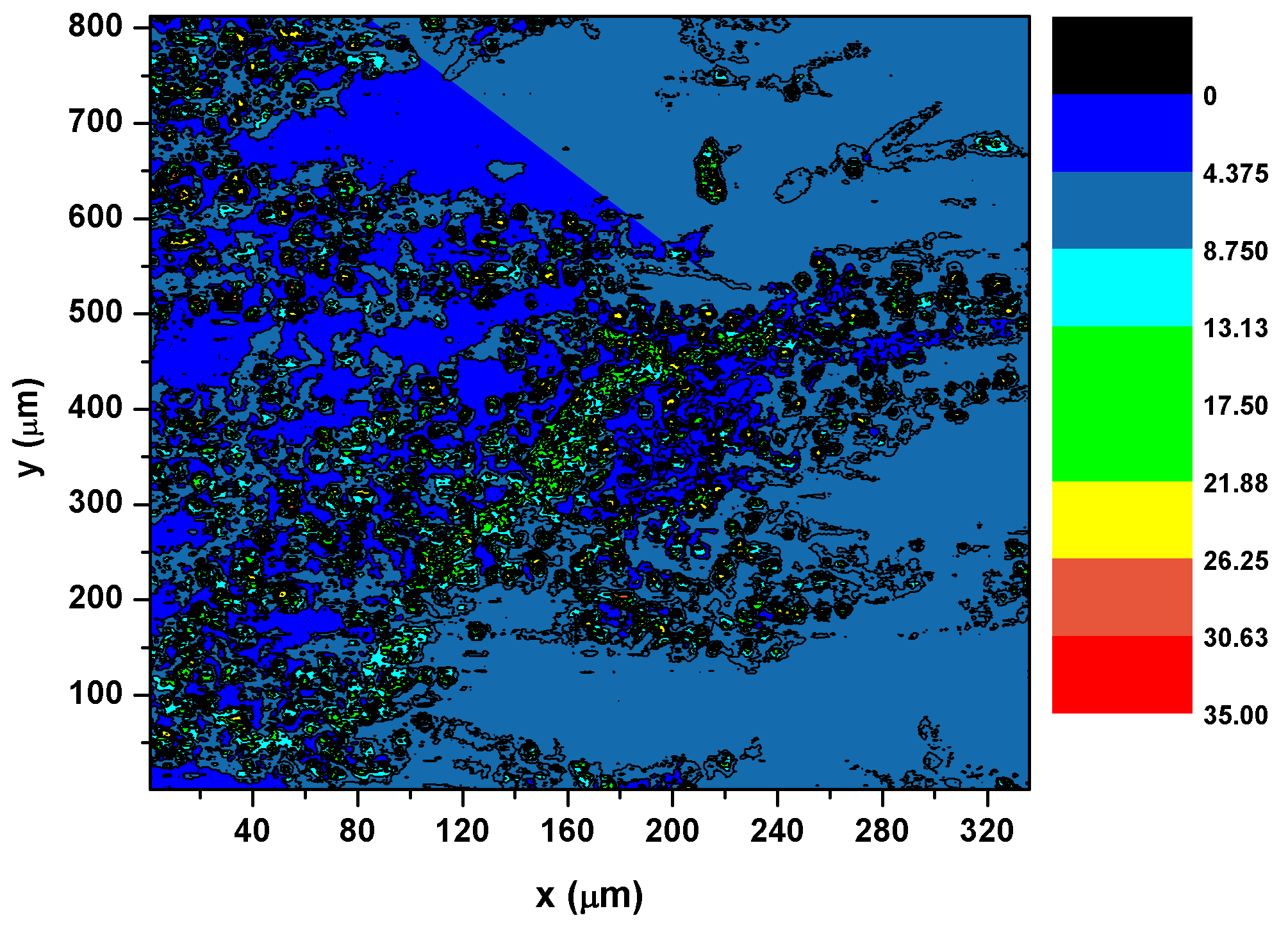}\label{aaT240}}\qquad
\subfloat[t = 240 min]{\includegraphics[scale=1.3]{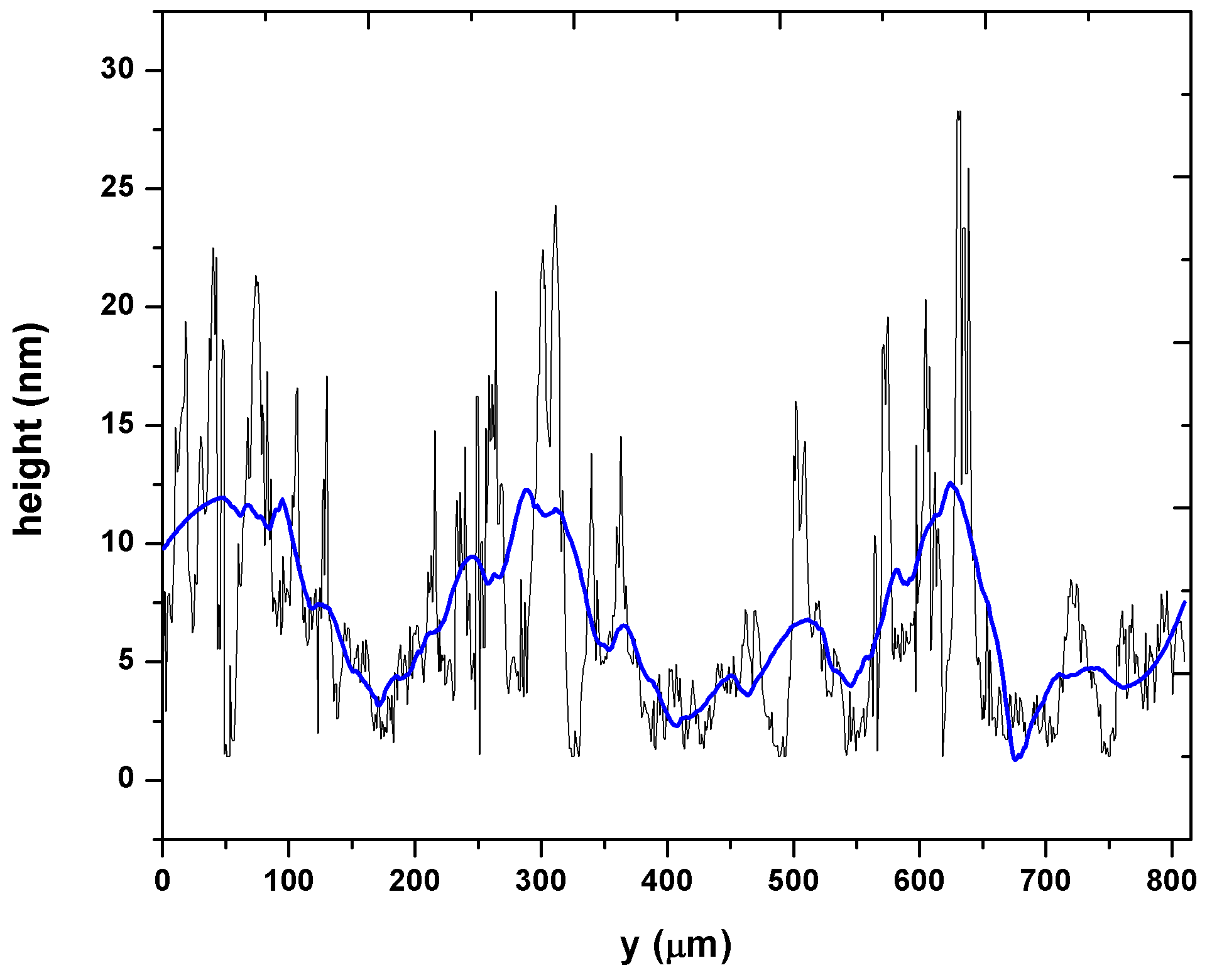}\label{lineT240}}
\caption{False colour height contour plots obtained from Imaging Ellipsometry of Arachidic Acid (C20) film on water transferred horizontally onto hydrophilic Si(100) substrates after the corresponding times after monolayer formation with the colour scale shown beside. A typical line profile from each contour plot follows the plot. The blue line is an average over 100 adjacent points.\label{contour}}
\end{figure}

Out-of-plane ND dynamics away from the collapse pressure of the monolayer is another important aspect of the long time scale behaviour. This requires the study of the height variation over a typical portion of the monolayer plane as a function of time. Height maps over a 337.00 $\mu m$  $\times$ 813.35 $\mu m$ (370 px $\times$ 893 px) window were extracted from the $\Delta$ maps provided by IE for C20 monolayers at $\pi =$ 25 mN/m and 25 $^\circ$C, deposited by the MILS technique on hydrophilic Si(100), as described in the Experimental section, at times of  30 min, 60 min, 90 min, 180 min and 240 min and shown in Figures \autoref{contour} \subref{aaT30}, \subref{aaT60}, \subref{aaT90}, \subref{aaT180}, \subref{aaT240} respectively in false colour with corresponding typical line profiles shown in Figures \autoref{contour} \subref{lineT30}, \subref{lineT60}, \subref{lineT90}, \subref{lineT180}, \subref{lineT240}. The thicker line in each of the latter represents an averaging over 100 adjacent points and gives the essectial feature of the height variation. They depict a complex dynamics that merits description.

\begin{figure}
\centering
\includegraphics[scale=2.0]{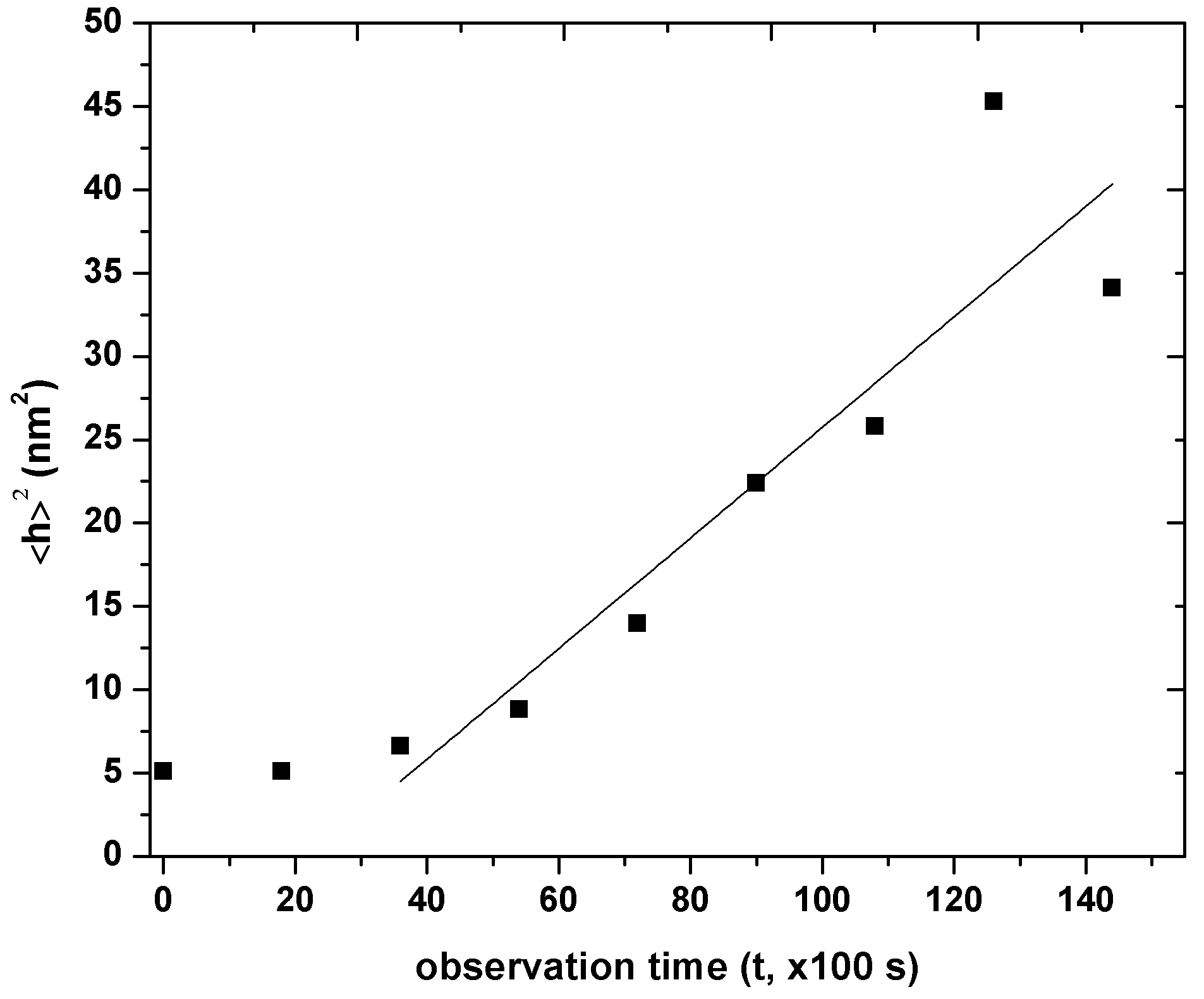}
\caption{Square of average height ($\langle h \rangle, nm$) of C20 films, transferred onto hydrophilic Si(100) and measured from Imaging Ellipsometry, plotted against  observation time (t,s). The linear fit of data is taken after 2000 s. }
\end{figure}

There are four major aspects of this dynamics -- growth of a flat or smooth film, growth of multilayered islands and growth of multilayer ridges from coalescence of these islands and evolution of `wave-like' features from coalescence of these ridges. The smooth film growth dominates till 60 min from the beginning and continues throughout but from the film at 60 min islands appear on this smooth surface and increase in both height and numbers, till, at 180 min onwards they cover the almost the entire surface. Ridges are observed at 120 min and they also grow in height and number with time. 

The nominal thickness of a C20 monolayer with untilted chains being $\approx$ 2.2 nm, we find that till 60 min there is a coexistence between a monolayer and a film of thickness between a bilayer and a trilayer, which, from free energy considerations we tentatively assign to a trilayer with the top layer having highly tilted chains. Till 30 mins the monolayer is dominant while from 60 min this trilayer starts to dominate till 180 min when a pentalayer (again with a tilted layer at top) appears and grows in coverage in the background.

The average height of islands grows from $\approx$ 4 nm to 20 nm. At 90 min we find some island clusters of size 30 $\mu m$ $\times$ 50 $\mu m$ but after that such isolated clusters are replaced by ridges spanning the field of view. The average height of the ridges grows from $\approx$  6 nm to 10 nm and their average thickness grows from 80 $\mu m$ to 250 $\mu m$ from 120 min to 240 min. They are found to grow roughly parallel to the trough barriers, i.e., perpendicular to the direction of shrinkage of monolayer area.  From 180 min onwards we see a new trend in the growth dynamics -- the emergence of wave-like structures, most probably from the lateral coalescence of ridges. They become prominant in Figure \autoref{contour} \subref{aaT240} and even more in the line profile of  \autoref{contour} \subref{aaT240} i.e., at 240 min. The height of these waves can reach 25-30 min while the `wavelength' is around 300 $\mu$m. Hence they resemble one-dimensional shallow waves on the surface of viscous fluid follwing Korteweg-de Vries equation \cite{korteweg1895} with the solution of Zabusky and Kruskal \cite{zabusky1965} which represents a set of solitons. Since shallow waves are generagted when the horizontal flow is much larger than the vertical flow and is smae at all depths of the liquid we propose that the in-plane coalescence of molecules is much faster that the upward diffusion and that this coalescence is the same at all levels of growth.
 
The square of the average height ($\langle h \rangle ^2$) of the film is plotted with time in Figure \label{diffusion}. The average height ($\langle h \rangle$) corresponds to the effective upward displancement of a molecule. From the plot it is clear that, after the first 2000 s ($\approx$ 30 min) when it remains constant (monolayer), $\langle h \rangle$ varies linearly with $t^{1/2}$, corresponding to upward diffusion of the moleucule. Thus the slope of the plot, calculated to be 0.0033 nm$^2$/s  gives the value of the upward diffusivity of the amphiphile.

Similar growth dynamics have been observed for monolayers of cobalt stearate, a two-tailed amphiphilic, at constant pressure collapse\cite{kundu2006}. The mechanism for the 2D to 3D transition is again diffusion as it was in that case. However, there are significant differences between the systems and the results of our studies and we shall enumerate them. (1) One major differences is, of course, that we are studying a monolayer at far below the collapse pressure of 42 mN/m whereas for the stearate monolayer the surface pressure went up to as high as 61 mN/m. Hence we are here in the stable zone of the monolayer in contrast to the completely unstable, collapsing monolayer in that other case. (2) the other difference is that our system is a monolayer of single-tailed amphiphiles while the stearate was, as told, a two-tailed amphiphile. We would expect a completely different behaviour in our system and the fact that the basic dynamics occurs through the formation of multtilayer -- both films and islands -- by upward diffusion of molecule in fatty-acids and stearate indicates an underlying universality in these 2D to 3D transitions. We find the growth to be in steps of bilayer and thus, along with the fact that diffusion through the monolayer and the subsequent multilayers can occur only if the hydrophilic nature of the carboxylic headgroups (due to the dipole moment)is suppressed, leads to the conclusion that adjacent fatty-acid molecules in the monolayer are dimerized at the headgroups through lipophilic attraction. Thus the lipophilic force again plays a pivotal role in the 2D to 3D transition and the dimer diffuses upwards with its two tails  dispossed symmetrically about the dimerized headgroups. This dimerization is, as discussed earlier, consistent with previous results obtained by Vysotsky\cite{vysotsky2012} and Goto\cite{goto2013}. Though both has the essential characteristics of Stranski-Krastanov (SK) growth, while the upward diffusion of molecules is present throughout the growth in this case of arachidic acid away from collapse, it could not be detected clearly in the collapse of cobalt stearate. However, the most important difference is the emergence of the set of `solitonic wave-like' structures from the in-plane coalescence of ridges on the film surface. The in-plane coalescence, which plays the dominant role in all of out-of-plane growth, is another evidence of the importance of the lipophilic force.

\section{Conclusion}
Mesoscopic (Brewster Angle Microscopy and Imaging Ellipsometry) and macroscopic (surface pressure-area or $\pi-A$ isotherm) methods have been used to understand the details of long term destabilization dynamics in Langmuir monolayers. We have used the isotherms to extract the monolayer area fraction versus time ($A_n-t$) curves and these form the bases of our macroscopic studies of the destabilization dynamics, while the BAM images form the mesoscopic bases. We have found significant differences in the results obtained from the two techniques. While the isotherm studies show the desorption dominated destabilization to be given by an exponentially decaying $A_n-t$ curve pointing to a single, continuous loss of the monolayer by dissolution, the BAM studies reveal a two-dimensional coalescence taking place simultaneously with the loss. On the other hand, a nucleation dominated destabilization emerges as a self-limiting process at both length scales. In both desorption and nucleation the initial step is a two-dimensional coalescence at the mesoscopic scale. Instabilities in monolayer are suggested to originate from packing defects at domain edges due to conflicting molecular orientations\cite{schief2000} or height differences at the boundary between 2D phases\cite{diamant2001} but our BAM studies show that 3D nucleation occurs in C20 and C18 monolayers with one single phase, consistent with Ybert et al\cite{ybert2002}.

From $A_n - t$ curves we have extracted the decay time constant ($\tau$) as a parameter to quantify the stability in Langmuir monolayers and looked at the effects of $\pi$ and temperature on $\tau$. This analysis shows that: (1) both desorption and nucleation are enhanced at higher surface pressures; for nucleation there exists a threshold pressure below which nucleation is absent whereas no threshold pressure is found for desorption and (2) both the destabilization mechanisms are enhanced with temperature. However, we have found that the most important factor regarding destabilization is the tail length of molecules. This molecular parameter not only decides the time constant of 2D-3D transformation on the external parameters but also causes a crossover from nucleation to desorption as tail length is decreased below a certain value keeping head group the same. Imaging Ellipsometry (IE) has been used to extract the height contour maps of the different stages of dynamics of out-of-plane growth of C20 monolayer, after horizontal transfer of the film onto Si(100) surface at these stages. It is seen that monolayer growth is followed successively by trilayer, multilayer islands, ridges through island-coalescence, and shallow wave-like structures through ridge-coalescence, having an essential resemblence with Stranski-Krastanov growth but with specific characteristics. While the molecules are transferred to the upper layers by diffusion, in-plane coalescence mediated by lipophilic attraction plays the crucial role in the evolution of these structure. Again, this attraction dimerizes the adjacent fatty-acid moelcules in the monolayer to initiate diffusion.

All these results have led us to conclude that the lipophilic attraction between the tails of the fatty acid monolayer is the driving or dominant force in the long term dynamics of fatty-acid Langmuir monolayers.

\section{Acknowledgement}
UKB thanks University Grants Commission (UGC) for their financial support and Director, Saha Institute of Nuclear Physics for giving permission to carry out research under the aegis of the Institute.

\bibliography{paper}

\end{document}